\documentclass[nofootinbib,prx,aps,superscriptaddress,notitlepage,reprint,preprintnumbers,showpacs,showkeys]{revtex4-1}
\usepackage{latexsym,graphicx,amssymb,amsmath,mathrsfs,color} 
\usepackage[utf8]{inputenc}
\usepackage{latexsym,graphicx,amssymb,amsmath}
\usepackage{setspace,bm}

\DeclareSymbolFont{bbold}{U}{bbold}{m}{n}
\DeclareSymbolFontAlphabet{\mathbbold}{bbold}

\newcommand{\be}{\begin{equation}}      
\newcommand{\ee}{\end{equation}}      
\newcommand{\bea}{\begin{eqnarray}}      
\newcommand{\eea}{\end{eqnarray}}    
     
\newcommand{\rt}[1]{{}}      
  
\newcommand{\Tr}{\,\textrm{Tr}\,}
\newcommand{\cc}{\,\textrm{c.c.}\,}

\newcommand{\rhs}{\,\textrm{rhs}\,} 
\newcommand{\lhs}{\,\textrm{lhs}\,} 
\newcommand{\sol}{\,\textrm{sol}\,}

\makeatletter
\renewcommand\appendix{\par
\setcounter{section}{0}
\setcounter{subsection}{0}
\gdef\thesection{\appendixname\space\@Alph\c@section}}

\long\def\unmarkedfootnote#1{{\long\def\@makefntext##1{##1}\footnotetext{#1}}}
\makeatother

\begin{document} 

\title{Vortex confinement transitions in the modified Goldstone model}

\author{Michikazu Kobayashi}
\email{michikaz@scphys.kyoto-u.ac.jp}
\affiliation{Department of Physics, Kyoto University, Oiwake-cho, Kitashirakawa, Sakyo-ku, Kyoto 606-8502, Japan}
\author{Gergely Fej\H{o}s}
\email{fejos@keio.jp}
\affiliation{Research and Education Center for Natural Sciences, Keio University, Hiyoshi 4-1-1, Yokohama, Kanagawa 223-8521, Japan}
\affiliation{Department of Physics, Keio University, Hiyoshi 3-14-1, Yokohama, Kanagawa, 223-8522, Japan}
\affiliation{Institute of Physics, E\"otv\"os University, 1117 Budapest, Hungary}
\author{Chandrasekhar Chatterjee}
\email{chandra.chttrj@gmail.com}
\affiliation{Research and Education Center for Natural Sciences, Keio University, Hiyoshi 4-1-1, Yokohama, Kanagawa 223-8521, Japan}
\author{Muneto Nitta}
\email{nitta@phys-h.keio.ac.jp}
\affiliation{Research and Education Center for Natural Sciences, Keio University, Hiyoshi 4-1-1, Yokohama, Kanagawa 223-8521, Japan}
\affiliation{Department of Physics, Keio University, Hiyoshi 4-1-1, Yokohama, Kanagawa 223-8521, Japan}

\begin{abstract}
{
The modified $XY$ model is a variation of the $XY$ model extended by a half periodic term, exhibiting a rich phase structure. 
As the Goldstone model,  also known as the linear O(2) model, can be obtained as a continuum and regular model for the $XY$ model, we define the modified Goldstone model as that of the modified $XY$ model.
We construct a vortex, a soliton (domain wall), and a molecule of two half-quantized vortices connected by a soliton as regular solutions of this model. 
Then we investigate its phase structure in two Euclidean dimensions 
via the functional renormalization group formalism and full numerical simulations. 
We argue that the field dependence of the wave function renormalization factor plays a crucial role in the existence of the line of fixed points describing the Berezinskii-Kosterlitz-Thouless (BKT) transition, which can ultimately terminate not only at one but at two end points in the modified model. This structure confirms that a two-step phase transition of the BKT and Ising types can occur in the system. We compare our renormalization group results with full numerical simulations, which also reveal that the phase transitions show a richer scenario than expected.
}
\end{abstract}

\maketitle

\section{Introduction}

The Berezinskii-Kosterlitz-Thouless (BKT) transition \cite{berezinskii71,berezinskii72,kosterlitz71,kosterlitz72} is 
a topological phase transition of two-dimensional systems, which divides a low-temperature phase with bound vortex-antivortex pairs from a high temperature phase with free vortices. The phenomenon was first analyzed in terms of the $XY$ model, and one of its most important impacts was that it showed that superfluidity and superconductivity can be realized even in two dimensions. Even though in two dimensions long-range order with continuous symmetry is forbidden by the Coleman-Mermin-Wagner (CMW) theorem \cite{Coleman:1973ci,mermin66,Hohenberg:1967zz}, there is a possibility of quasi-long-range order, which shows algebraically decaying correlations. The BKT transition realizes this scenario, and it also has the unique feature of being a continuous phase transition without breaking any symmetry.  
It has been experimentally confirmed in various condensed matter systems 
such as 
$^4$He films \cite{Bishop}, thin superconductors \cite{Gubser,Hebard,Voss,Wolf,Epstein}, Josephson-junction arrays \cite{Resnick,Voss2}, colloidal crystals \cite{Halperin,Young,Zahn,Nakamura}, and ultracold atomic Bose gases \cite{Hadzibabic}. 
The $XY$ model shares common properties including the BKT transition 
with the two-dimensional linear O(2), or Goldstone model
at large distances or low energies, 
which is a regular version of the $XY$ model described by one complex scalar field,
in which the U(1) Goldstone mode for the $XY$ model 
is complemented by a massive amplitude (Higgs) mode.
One of the merits of the latter is to allow vortices as regular solutions
in contrast to the $XY$ model in which vortices are singular configurations.

$XY$-like models do not necessarily show the BKT transition. For example, for sharply increasing spin-spin potential, the phase transition between the paramagnetic and ferromagnetic phases can be of first order \cite{domany84}. It is not surprising that the so-called modified $XY$ model, where on a square lattice  the Hamiltonian of the rotor is extended with a $\pi$ periodic term
\bea
\label{Eq:modXY}
{\cal H}_{\rm mXY}=-J\sum_{\langle i,j \rangle} \cos (\vartheta_i-\vartheta_j)-J' \sum_{\langle i,j \rangle} \cos [2(\vartheta_i-\vartheta_j)], \nonumber\\
\eea
also shows a different scenario. It was predicted long ago that for large enough $J'$ coupling, there exists a nematic phase separated from the ferromagnet and the transition between them is of Ising type  \cite{korshunov85,lee85}. 
This was also confirmed by numerical calculations \cite{carpenter89}. 
The Ising-type transition is related to the presence of domain walls in this model. Moreover,
it was conjectured that molecules and anti molecules of half-quantized vortices play a crucial role for phase transitions,  
in contrast to a pair of vortices and anti vortices in the $XY$ model. 
As of today, the model (\ref{Eq:modXY}) and its various modifications \cite{dian11,shi11,bonnes12,huebscher13,serna17,nui18,canova16,zukovic17,zukovic18} are of great importance and interest, especially due to their relevance in condensed matter physics applications, e.g., superfluidity in atomic Bose gases \cite{radzihovsky08}, arrays of unconventional Josephson junctions \cite{korshunov10}, or high-temperature superconductivity \cite{komendova10}. 

The BKT transition of the $XY$ model was originally analyzed via a real-space renormalization group (RG) approach \cite{kosterlitz72}, which is rather unconventional and not easily linkable to the Wilsonian picture of the RG \cite{herbut}. In the past, the functional RG (FRG) approach, which adopts the Wilsonian idea of mode elimination and averaging to the level of the effective action \cite{kopietz}, was also applied and developed in regard to the BKT transition in both continuum \cite{grater95,gersdorff01,jakubczyk14,jakubczyk17,defenu17} and lattice formulations \cite{machado10,krieg17}. It turned out that the conventional, Wetterich formulation of the method was capable of showing signs in the two-dimensional linear O(2) or Goldstone model of the line of fixed points that is responsible for the topological nature of the phase transition. This is remarkable in the sense that no vortices need to be introduced explicitly, as opposed to the older real-space RG description \cite{kosterlitz72}. One of the shortcomings of the treatment, however, is that because one is typically solving the RG flow equation of the scale-dependent effective average action via a derivative expansion, as an artifact, only a line of quasi-fixed points is found. That is, the RG flow does not stop along this line, but only slows down significantly compared to other regions of the parameter space. It is worth pointing out that recently in a dual lattice formulation of the FRG, Krieg and Kopietz \cite{krieg17} exactly reproduced the RG flow equations derived by Kosterlitz and Thouless \cite{kosterlitz72} and therefore the existence of a true line of fixed points was established in terms of a momentum space RG. It would be interesting, however, to develop a scheme in the ordinary Wetterich formulation of the FRG, which could also lead to a similar result.

The goal of this study is twofold. On the one hand, we aim to show a rather simple approximation scheme of the FRG flow equations that can show significant improvement on the possibility of reaching a true line of fixed points in the continuum version of the $XY$ model, and more importantly argue that it can also be applied naturally to the modified $XY$ model, i.e., the continuum version of (\ref{Eq:modXY}). In the framework of a momentum space RG, we describe the two-step transition in the latter model and we will also predict that fluctuations may completely make the topological transition disappear. On the other hand, we also aim to provide full numerical simulation of the system and show that depending on the value of the self-coupling of the scalar field, the structure of the transitions is even richer than it is predicted by the RG.

The paper is organized as follows. In Sec.~\ref{Sec:basics}
we introduce the modified Goldstone model and construct classical solutions, 
an integer vortex, a soliton, and a vortex molecule of two half-integer vortices connected by a soliton in that model.
 In Sec.~\ref{sec:FRG}, 
 after giving a brief review of the FRG, we reproduce some earlier results of the BKT transition via the FRG and also show the improvement announced above. Then this scheme is applied to the modified $XY$ model and we show how a two-step transition can emerge in the system. In Sec.~\ref{sec:numerics} we confirm this scenario via full numerical simulations and reveal the nature of the corresponding transitions. Section ~\ref{sec:summary} is devoted to a summary.
In Appendix A
we show how to derive the Hamiltonian of the modified Goldstone model from the microscopic lattice model of the modified $XY$ model, while in Appendix B
we derive some of the corresponding flow equations of the FRG.

\begin{figure}
\centering
\includegraphics[width=.71\linewidth]{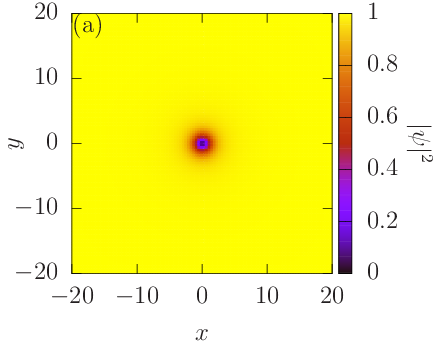} \\[5pt]
\includegraphics[width=.71\linewidth]{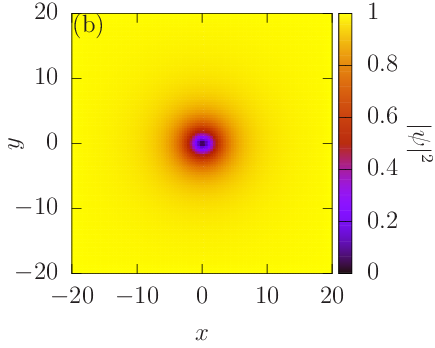} \\[5pt]
\includegraphics[width=.71\linewidth]{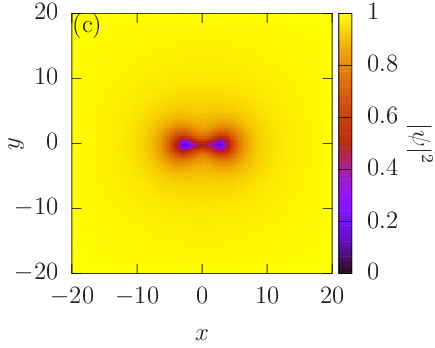} \\[5pt]
\includegraphics[width=.71\linewidth]{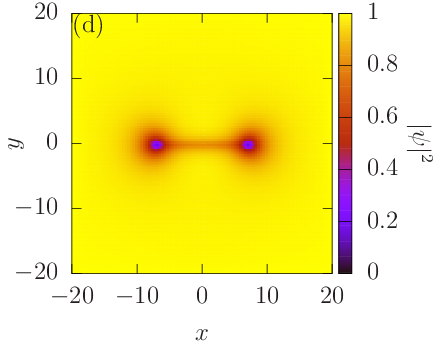}
\caption{Vortex solution of the field equations for $\lambda = 8$, $\rho_0 = 1/2$, and (a) $\theta = 0^\circ$, (b) $\theta = 77^\circ$, (c) $\theta = 78^\circ$, (c) $\lambda = 85^\circ$.
It transforms into a half-quantized vortex molecule around $\theta \approx 78^\circ$.}
\label{Fig1}
\end{figure}

\section{Model and Solutions}
\label{Sec:basics}

\subsection{Modified Goldstone model}

In this study we are interested in the continuum version of the $XY$ model, i.e., the Goldstone model and its modification [for its derivation from the microscopic Hamiltonian (\ref{Eq:modXY}) see Appendix A]
\bea
\label{Eq:Ham}
{\cal H}&=&\int_x \left[ a|\nabla \psi|^2 +b|\nabla \psi^2|^2+\frac{\lambda}{2}\big(|\psi|^2/2-\rho_0\big)^2\right],
\eea
where $\psi$ is a complex scalar field, and $\lambda$, $a$, and $b$ are positive coupling constants. The continuum version of the standard $XY$ model refers to $b=0$ and in the modified $XY$ model we have $b> 0$. The field equation can be obtained from 
the Hamiltonian (\ref{Eq:Ham}) as
\bea
\!\!\!0 = \frac{\delta \mathcal{H}}{\delta \psi^\ast} = - a\, \Delta \psi - 2 b\, \psi^\ast\, \Delta\psi^2 + \frac{\lambda}{2} \left(\frac{|\psi|^2}{2} - \rho_0 \right) \psi,
\label{Eq:modified-GP}
\eea
which we call the modified Gross-Pitaevskii equation.

\subsection{Classical solutions}

Field equations (\ref{Eq:modified-GP}) 
of the modified Goldstone model admit superfluid (or global) vortex solutions.
%Before we analyze the renormalization group flows and perform a full numerical simulation of the thermal behavior of the system, 
Here we show how such a vortex solution transforms into a half-quantized vortex molecule, when the second term of Eq.~(\ref{Eq:Ham}) becomes large enough. As we wish to compare our results with earlier works \cite{carpenter89}, in what follows we work in a simplified parameter space, where $a^2+b^2=1$, and thus the $a=\cos\theta$ and $b=\sin\theta$ parametrization can be used. As it turns out, this choice also helps perform the full numerical simulations of the thermodynamics of the system more easily. The transformation of the vortex solution can be seen in Fig.~\ref{Fig1}. One observes that around $\theta \approx 78^\circ$, a clear picture of a vortex molecule emerges, where two half-quantized vortices are connected by a one-dimensional soliton. One expects that at finite temperature, as a function of $\theta$, somewhere close to the aforementioned value, the emergence of the molecules will have an effect on the phase structure of the system.

\if0 %%%
:
\begin{align}
\begin{split}
&\quad \mathcal{H}_{\rm 1D}= \\
& \int dx\: \left[ a|\nabla \psi|^2 +b|\nabla \psi^2|^2 + \frac{\lambda}{2}\left(|\psi|^2/2-\rho_0\right)^2\right], \\
& 0 = \frac{\delta \mathcal{H}_{\rm 1D}}{\delta \psi^\ast} \\
& \phantom{0} = - a\, \Delta \psi - 2 b\, \psi^\ast\, \Delta\psi^2 + \frac{\lambda}{2} \left(\frac{|\psi|^2}{2} - \rho_0 \right) \psi,
\end{split}
\label{Eq:modified-GP}
\end{align}
\fi  %%%
%%%%%%%%%%%%%%%%%%%%%%
\begin{figure}[htb]
\centering
\includegraphics[width=0.6\linewidth]{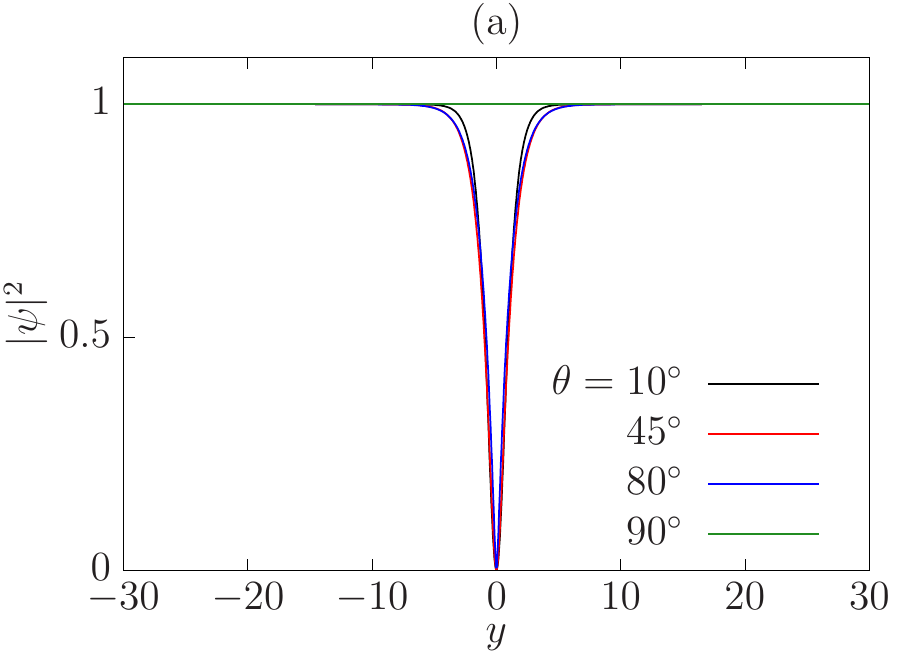} \\[10pt]
\begin{minipage}{0.494\linewidth}
\centering
\includegraphics[width=0.99\linewidth]{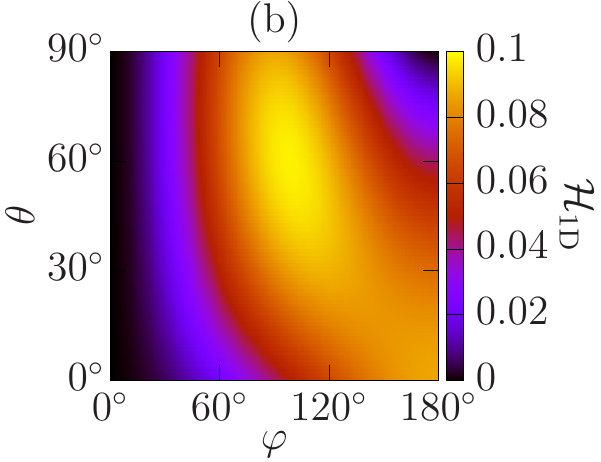}
\end{minipage}
\begin{minipage}{0.494\linewidth}
\centering
\includegraphics[width=0.99\linewidth]{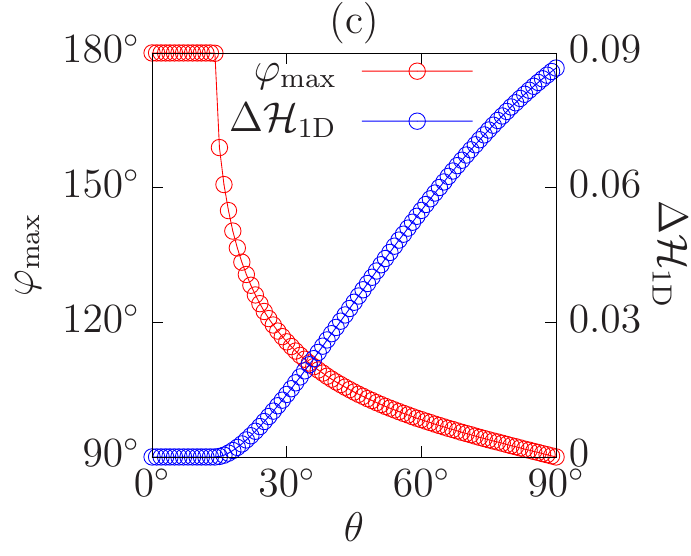}
\end{minipage}
\caption{
\label{Fig:soliton-energy}
(a) Profiles of the amplitude $|\psi|^2$ of the soliton solutions for the modified Gross-Pitaevskii equation [Eq. \eqref{Eq:modified-GP}] with $\theta = 10^\circ$ (black), $\theta = 45^\circ$ (red), $\theta = 80^\circ$ (blue), and $\theta = 90^\circ$ (green).
(b) Dependence of the energy, $\mathcal{H}_{\rm 1D}$, on $\theta$ and $\varphi$.
(c) Dependence of the maximal angle $\varphi_{\rm max}$ and the energy barrier $\Delta \mathcal{H}_{\rm 1D}$ on $\theta$.
In the both panels, we set $\lambda = 8$ and $\rho_0 = 1 / 2$.
}
\end{figure}
%%%%%%%%%%%%%%%%%%%%

In a vortex molecule shown in Fig.~\ref{Fig1}, each of the two vortices has a half-quantized circulation $\int d\vec{l}  (\vec{\nabla} \mathrm{arg}[\psi]) = \pi$ and the soliton connecting them has a $\pi$-phase jump. To analyze the stability of the soliton, we determine the following one-dimensional stable solution of the modified
Gross-Pitaevskii equation (\ref{Eq:modified-GP}) in one dimension with the boundary condition $\psi(y \to -\infty) = \sqrt{2 \rho_0}$, and $\psi(y \to \infty) = \sqrt{2 \rho_0} e^{i \varphi}$.
Fig.~\ref{Fig:soliton-energy} (a) shows the profiles of the soliton solutions, while
Fig.~\ref{Fig:soliton-energy} (b) shows the total energy $\mathcal{H}_{\rm 1D}$ as a function of $\varphi$ and $\theta$.
It is clear that if $\mathcal{H}_{\rm 1D}$ takes the maximum value at some $\varphi < \pi$, then the soliton solution with $\varphi = \pi$ becomes 
locally stable (metastable) by having a positive energy barrier $\Delta \mathcal{H}_{\rm 1D} \equiv \mathcal{H}_{\rm 1D}(\varphi = \varphi_{\rm max}) - \mathcal{H}_{\rm 1D}(\varphi = \pi)$, where the maximal angle $\varphi_{\rm max}$ is the value of $\varphi$ at which $\mathcal{H}_{\rm 1D}$ takes the maximum.
Fig.~\ref{Fig:soliton-energy} (c) shows the maximal angle $\varphi_{\rm max}$ and the energy barrier $\Delta \mathcal{H}_{\rm 1D}$.
The former starts to take a nonzero value, $\Delta \mathcal{H}_{\rm 1D} > 0$, with $\varphi_{\rm max} < \pi$ at around $\theta \approx 15^\circ$, above which the soliton is, therefore, energetically stable. That is to say, the appearance of vortex molecules and the stability of the soliton are not related, and thus it is not the (de)stabilization of the domain wall that lets molecules emerge. 

It is worth noting that these configurations become singular in the limit of $\lambda \to \infty$, in which the model reduces to the modified $XY$ model. Therefore, the modified $XY$ model does {\it not} allow these configurations as solutions to the field equations, while the modified Goldstone model does.

\subsection{Type of symmetry and (quasi) breaking of symmetry}

Here, we discuss the symmetry properties of the Hamiltonian 
[Eq.~\eqref{Eq:Ham}] and show the possible 
(quasi-)breaking patterns of symmetries.
The symmetry of the Hamiltonian with generic parameters is of $\mathrm{U}(1)$ as a phase shift of the field, $\psi \to \psi e^{i \alpha}$ for the arbitrary $\alpha \in [0, 2\pi)$.
In the case of $a = 0$ and $b > 0$, the two fields $\psi$ and $\psi\, e^{i \pi}$ are identifiable, because the Hamiltonian [Eq. \eqref{Eq:Ham}] is the functional of $\psi^2$ rather than $\psi$.
Therefore, the symmetry of the Hamiltonian is only $\mathrm{U}(1) / \mathbb{Z}_2$, where the $\mathbb{Z}_2$ symmetry comes from the identification of $\psi \sim \psi\, e^{i \pi}$.
This $\mathbb{Z}_2$ factor is essential for the presence of (deconfined) half-quantized vortices.

Depending on the parameter regions, 
the $\mathrm{U}(1)$ or $\mathrm{U}(1)/\mathbb{Z}_2$ symmetry is spontaneously broken in the ground state in different patterns summarized as follows: 
\begin{subequations}
\label{eq:ssb}
\begin{align}
&\mathrm{U}(1) \stackrel{\mathrm{U}(1)}{\dashrightarrow} 1 \qquad && \text{for $a > 0$ and $b = 0$}, \label{eq:ssb-simple-BKT} \\
&\mathrm{U}(1)/\mathbb{Z}_2 \stackrel{\mathrm{U}(1)/\mathbb{Z}_2}{\dashrightarrow} 1 \qquad && \text{for $a = 0$ and $b > 0$}, \label{eq:ssb-half-BKT} \\
&\mathrm{U}(1) \stackrel{\mathrm{U}(1)/\mathbb{Z}_2}{\dashrightarrow} \mathbb{Z}_2 \stackrel{\mathbb{Z}_2}{\longrightarrow} 1 \qquad && \text{for $b \gg a > 0$}, \label{eq:ssb-two-step} \\
&\mathrm{U}(1) \stackrel{\mathrm{U}(1)}{\Longrightarrow} 1 \qquad && \text{for $a \approx b$} \label{eq:ssb-BKT-TT}.
\end{align}
\end{subequations}
Here the arrows $\dashrightarrow$, $\longrightarrow$, and $\Longrightarrow$ 
denote quasi-breaking of symmetry via a BKT transition, ordinary symmetry breaking with a thermodynamic phase transition, and simultaneous 
(quasi)breaking of symmetry, respectively. Here, quasibreaking means that the symmetry is not exactly broken due to the CMW theorem in the thermodynamic limit but is locally broken at semi macroscopic scales with an algebraically decaying correlation function. 

Now let us explain each breaking pattern.
In the simplest case, i.e., for $a > 0$ and $b = 0$  [Eq. \eqref{eq:ssb-simple-BKT}], the standard BKT transition occurs with the quasibreaking of the $\mathrm{U}(1)$ symmetry.
In the opposite case, i.e., for $a = 0$ and $b > 0$  [Eq. \eqref{eq:ssb-half-BKT}], 
the BKT transition occurs with the quasi breaking of the $\mathrm{U}(1) / \mathbb{Z}_2$ symmetry, for which half-quantized and anti-half-quantized vortices start to form in pairs.
In the case of $b \gg a > 0$  [Eq. \eqref{eq:ssb-two-step}], 
two successive spontaneous (quasi)breaking processes occur. 
At the first stage (at higher temperature)
the $\mathrm{U}(1)$ symmetry is quasi broken to 
a $\mathbb{Z}_2$ subgroup 
accompanied by the BKT transition. 
At the second stage, at
a temperature lower than the BKT transition temperature, the remaining $\mathbb{Z}_2$ symmetry is further spontaneously broken due to a thermodynamic transition.
In this case, half-quantized and anti-half-quantized vortices start to form pairs at the BKT transition and domain walls appear at the thermodynamic transition.
Some domain walls have no endpoint forming loops as well as those in the Ising model, but some others appear between two half-quantized or two anti-half-quantized vortices forming vortex or anti-vortex molecules as shown in Fig. ~\ref{Fig1}.
In the remaining case of $a \approx b$  [Eq.~(\ref{eq:ssb-BKT-TT})], rather than a conventional BKT transition, the BKT transition occurs with the quasibreaking of $\mathrm{U}(1) / \mathbb{Z}_2$ symmetry and the thermodynamic transition with breaking of $\mathbb{Z}_2$ symmetry simultaneously.
All vortices are integers and domain walls do not have endpoints.

In the following sections, we study the modified Goldstone model 
by the FRG and Monte Carlo simulation.

%%%%%%%%%%%%%%%%%%%%%%%%%%%%%%%%%%
\section{Functional renormalization group calculations}\label{sec:FRG}

In this section, after giving a brief review of FRG, 
we apply it to the modified Goldstone model approximately, at the leading order of the derivative expansion, and obtain the phase structure.

\subsection{Flow equation: a review}

Here we review the basics of the FRG. At the core of the formalism lies the $\Gamma_k$ average effective action, in which fluctuations of the dynamical fields are incorporated up to a momentum scale $k$. The $\Gamma_k$ function obeys the flow equation:
\bea
\label{Eq:flow1}
\partial_k \Gamma_k = \frac12 \int \Tr [(\Gamma_k^{(2)}+R_k)^{-1}\partial_k R_k],
\eea
where $\Gamma_k^{(2)}$ is the second derivative matrix of $\Gamma_k$ with respect to the dynamical variables and $R_k$ is a regulator function, which is defined (in Fourier space) through a momentum-dependent mass term
\bea
\frac12 \int_{p,q} \psi^i (q) R^{ij}_k (q,p) \psi^j (p),
\eea
added to the classical Hamiltonian (or Euclidean action). We denoted the set of fluctuating field variables by $\psi$. Here $R_k$ is supposed to give a large mass to modes that have momenta $q \lesssim k$ and leave the ones with $q\gtrsim k$ untouched. The classical Hamiltonian by definition does not contain any fluctuations; therefore, it serves as an initial condition for the RG flow of $\Gamma_{k=\Lambda}$ at some microscopic scale $\Lambda$. The flow equation (\ref{Eq:flow1}) then needs to be integrated down to $k=0$, where one obtains the full free energy (or quantum effective action). One is free to choose the $R_k$ function such that it fulfills the requirement of suppressing low-momentum modes, and in this paper we employ the so-called optimal version:
\bea
\label{Eq:optreg}
R_k(q,p)=Z_k(2\pi)^2 (k^2-q^2)\Theta(k^2-q^2) \delta(q+p),
\eea
where $\Theta(x)$ is the Heaviside step function, and $Z_k$ is the wave function renormalization factor.

\subsection{Local potential approximation'}
Here we solve flow equation (\ref{Eq:flow1}) for the modified Goldstone model approximately, using the  ansatz for $\Gamma_k$,
\bea
\label{Eq:LPAp}
\Gamma_k = \int d^2x \left[ \frac{Z_k(\rho)}{2} (\nabla \psi^i)^2 + \frac{\lambda_k}{2}(\rho-\rho_{0,k})^2\right],
\eea
where instead of a complex variable, the $\psi^i$ field is considered as a two-component real vector: $\psi^i=(\psi^1, \psi^2)$, while $\rho=\psi^i\psi^i/2$, and we have only kept the original couplings in the effective potential. Namely, Eq.~(\ref{Eq:LPAp}) is compatible with the form of Eq.~(\ref{Eq:Ham}), but it comes with $k$-dependent couplings and a field-dependent wave function renormalization factor [$Z_k(\rho)$]. In what follows we will consider the $Z_k(\rho)$ function in two separate approximations: $i)$ $Z_k(\rho)\approx Z_k(\rho_0)$, and $ii)$ $ Z_k(\rho)\approx Z_k(\rho_0)+Z_k'(\rho_0)(\rho-\rho_0)$. Approximation i) is sometimes called the local potential approximation' (LPA'), with the prime referring to nontrivial wave function renormalization. First we work with the LPA' and the next section is devoted to approximation $ii)$.

 Projecting the flow equation (\ref{Eq:flow1}) onto a subspace spanned by homogeneous field configurations, we get (see also Appendix B)

\begin{figure*}
\begin{center}
\raisebox{0.05cm}{
\includegraphics[bb = 100 0 255 200,scale=0.65,angle=0]{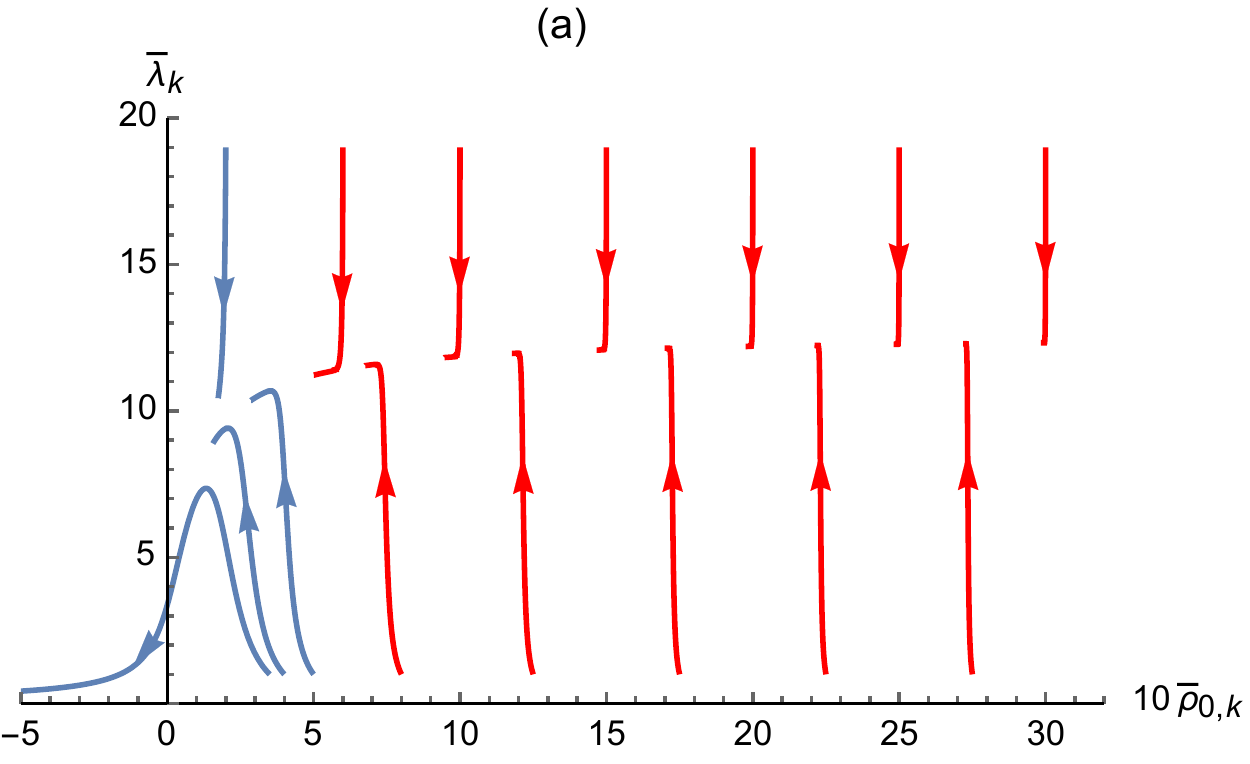}}
\includegraphics[bb = -140 0 255 200,scale=0.65,angle=0]{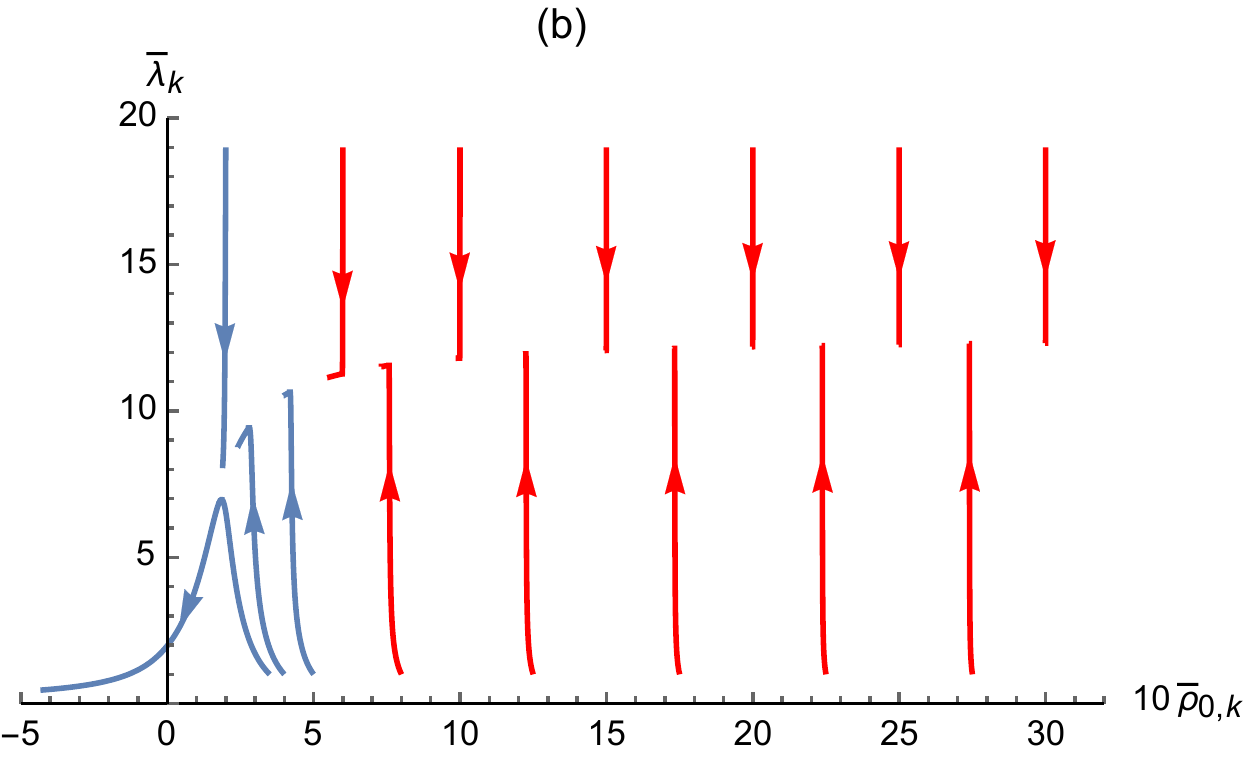}
\caption{Comparison of (a) the leading order flow diagram with (b) the wave function renormalization improved one. The red flows are stopped at (a) $t=-\log(k/\Lambda)=10$ and (b) $t=200$ (b), which shows that a significant stabilization of the line of fixed points is achieved with the improved approximation. For completeness, the blue flows are stopped at (a) $t=1,5,5,8$ and (b) $t=2,8.5,20,40$ (b), respectively.}
\label{Fig:flows}
\end{center}
\end{figure*}

\begin{subequations}
\label{Eq:lambdarhoflow}
\bea
\label{Eq:lambdaflow}
k\partial_k \bar{\lambda}_k= -2\bar{\lambda}_k [1 - \eta^{(0)}_k]&+&\frac{\bar{\lambda}_k^2}{2\pi} \left(1-\frac{\eta_k^{(0)}}{4}\right) \nonumber\\
&\times& \left[1+\frac{9}{(1+2\bar{\rho}_{0,k}\bar{\lambda}_k)^3}\right], \\
\label{Eq:rhoflow}
k\partial_k \bar{\rho}_{0,k}=-\eta^{(0)}_k\bar{\rho}_{0,k}&+&\frac{1}{4\pi}\left(1-\frac{\eta^{(0)}_k}{4}\right) \nonumber\\
&\times&\left[1+\frac{3}{(1+2\bar{\rho}_{0,k}\bar{\lambda}_k)^2}\right],
\eea
\end{subequations}
where we have introduced dimensionless rescaled variables $\bar{\lambda}_k=\lambda_k k^{-2}Z_k^{-2}$ and $\bar{\rho}_{0,k}=\rho_{0,k}Z_k$. Here, $\eta^{(0)}_k=-k\partial_k Z_k/Z_k$ is the anomalous dimension at this order of the approximation, where the wave function renormalization is evaluated at the minimum point of the effective potential, $Z_k\equiv Z_k(\bar{\rho}_{0,k})$ (from now on we think of the wave function renormalization as a function of the rescaled field). If we project Eq.~(\ref{Eq:flow1}) onto $\sim (\nabla \psi^t)^2$, where the index refers to the transverse direction, we arrive at the flow equation for $Z_k$ (see, again, Appendix B for details),
\bea
\label{Eq:Zkflow}
k\partial_k Z_k(\bar{\rho})= -Z_k(\bar{\rho}) \frac{\bar{\rho}\bar{\lambda}_k^2/\pi}{(1+\bar{M}_{l,k}^2)^2(1+\bar{M}_{t,k}^2)^2}, 
\eea
where $\bar{M}_{l,k}^2=M_{l,k}^2/Z_kk^2$ and $\bar{M}_{t,k}^2=M_{t,k}^2/Z_kk^2$, while $M_{l,k}^2$ and $M_{t,k}^2$ are the longitudinal and transverse components of the momentum independent part of the $\Gamma_k^{(2)}$ matrix, respectively,
\bea
M_{l,k}^2=\lambda_k(3\rho-\rho_{0,k}), \quad M_{t,k}^2=\lambda_k(\rho-\rho_{0,k}),
\eea
and thus
\bea
\bar{M}_{l,k}^2=\bar{\lambda}_k(3\bar{\rho}-\bar{\rho}_{0,k}), \quad \bar{M}_{t,k}^2=\bar{\lambda}_k(\bar{\rho}-\bar{\rho}_{0,k}).
\eea
Since in Eqs.~(\ref{Eq:lambdarhoflow}) it is $Z_k=Z_k(\bar{\rho}_{0,k})$ that appears through $\eta_k^{(0)}$, we evaluate Eq.~(\ref{Eq:Zkflow}) at $\bar{\rho}=\bar{\rho}_{0,k}$ and get
\bea
\label{Eq:eta0flow}
\eta^{(0)}_k = \frac{\bar{\rho}_{0,k}\bar{\lambda}_k^2}{\pi(1+2\bar{\rho}_{0,k}\bar{\lambda}_k)^2}.
\eea
Now we can search for fixed points of Eqs.~(\ref{Eq:lambdarhoflow}) and (\ref{Eq:eta0flow}). The flow diagram in terms of $\bar{\lambda}_k$ and $\bar{\rho}_{0,k}$ can be seen on the left side of Fig.~\ref{Fig:flows}. We observe the line of quasifixed points and notes that the flow, even though significantly slowed down, is clearly nonzero in the aforementioned region.

\subsection{Wave function renormalization improvement}

The key to the improvement to be described here is to realize how crucial the role of the wave function renormalization factor $Z_k$ is in the previous description. In order to escape from the CMW theorem, in the low-temperature phase $Z_k$ has to diverge so that the renormalized field can condense (the expectation value of the bare field is always zero). Since any rescaling of the field should lead to the same description of the system, one expects that any field derivative of the wave function renormalization factor is proportional to $Z_k$ itself, $Z^{(n)}_k \sim Z_k$, which means that they also diverge, and in principle none of them should be neglected,  as also pointed out, e.g., in \cite{jakubczyk14}. As announced in the preceding section, here we take into account the first derivative of $Z_k$, which will indeed lead to a significant improvement in stabilizing the flow along the (quasi)line of fixed points, but more importantly, it also makes it possible to treat the modified Goldstone model in the FRG.

If we keep track of the field derivative of $Z_k$, then it is possible to take into account in $\eta_k$ the implicit $k$ dependence coming from the change of the minimum of the effective potential when the RG scale is varied, similarly to what was done in earlier works, e.g., \cite{tetradis94,rose17}. In principle, we should have
\bea
\eta_k &=& -\frac{kd_k Z_k}{Z_k}=-\frac{k\partial_k Z_k+Z_k'k\partial_k \bar{\rho}_{0,k}}{Z_k} \nonumber\\
&=&\eta_k^{(0)}-w_kk\partial_k\bar{\rho}_{0,k}\equiv \eta_k^{(0)}+\Delta \eta_k,
\eea
where $d_k$ refers to total differentiation and on the right-hand side both $Z_k$ and $Z_k'$ are evaluated at $\bar{\rho}=\bar{\rho}_{0,k}$. We have also introduced the notation $\Delta \eta_k=-w_k k\partial_k \bar{\rho}_{0,k}$ with $w_k=Z_k'/Z_k$. Since $Z_k'$ has appeared in our formula, we also need to derive a flow equation for it. This can be obtained from Eq.~(\ref{Eq:Zkflow}) after applying $d/d\bar{\rho}$ to the both sides (note that $\partial_k$ does not commute with $d/d\bar{\rho}$). Using that $d\bar{M}^2_{l,k}/d\bar{\rho}=3\bar{\lambda}_k$ and $d\bar{M}^2_{t,k}/d\bar{\rho}=\bar{\lambda}_k$, we get
\bea
\label{Eq:Zprimeflow}
k\partial_k Z_k'(\bar{\rho}_{0,k})&&/Z_k(\bar{\rho}_{0,k})=\\
&&\frac{4\bar{\rho}_{0,k}^2\bar{\lambda}_k^2+6\bar{\rho}_{0,k}\bar{\lambda}_k+2\bar{\rho}_{0,k}w_k-1}{\pi(1+2\bar{\lambda}_k\bar{\rho}_{0,k})^3}+w_k\eta_k^{(0)},\nonumber
\eea
which leads to 
\bea
\label{Eq:wkflow}
k\partial_k w_k &=&\frac{4\bar{\rho}_{0,k}^2\bar{\lambda}_k^2+6\bar{\rho}_{0,k}\bar{\lambda}_k+2\bar{\rho}_{0,k}w_k-1}{\pi(1+2\bar{\lambda}_k\bar{\rho}_{0,k})^3}\nonumber\\
&+&2w_k\eta_k^{(0)}-w_k^2k\partial_k\bar{\rho}_{0,k}.
\eea
At this point, it is important to mention that Eq.~(\ref{Eq:Zprimeflow}) is not exact, as deriving Eq.~(\ref{Eq:Zkflow}) we let the field operators act only on the potential part of the two-point correlation function and not on $Z_k(\rho)$. This would have introduced a further $Z_k'(\rho)$ dependence on the right-hand side of Eq.~(\ref{Eq:Zkflow}), which is neglected here. The reason behind this is that we think of the scheme in question as a first correction to the LPA' in the sense that we are motivated to derive the flow of $w_k$ in the background flows of $Z_k$, $\bar{\lambda}_k$, and $\bar{\rho}_{0,k}$ of the LPA', which by definition are not affected explicitly by $w_k$ itself.

Now, once we return to the aforementioned flows, i.e., Eqs. (\ref{Eq:lambdarhoflow}), we notice that they do depend implicitly on $w_k$, but only because of the new expression of the anomalous dimension $\eta_k^{(0)} \rightarrow \eta_k= \eta_k^{(0)}+\Delta \eta_k$. This does not make much of a difference in the flow of $\bar{\lambda}_k$, but changes that of $\bar{\rho}_{0,k}$. The reason is that Eq.~(\ref{Eq:rhoflow}) becomes an implicit equation, since $k\partial_k \bar{\rho}_{0,k}$ also appears on the right-hand side through $\Delta \eta_k \equiv -w_kk\partial_k \bar{\rho}_{0,k}$. After some algebra we arrive at
\bea
\label{Eq:rhoflow2}
k\partial_k \bar{\rho}_{0,k}=\frac{-\eta_k^{(0)}\bar{\rho}_{0,k}+\frac{1}{4\pi}\left(1-\frac{\eta_k^{(0)}}{4}\right)\left[1+\frac{3}{(1+2\bar{\rho}_{0,k}\bar{\lambda}_k)^2}\right]}{1-w_k\left[\bar{\rho}_{0,k}+\frac{1}{16\pi}\left(1+\frac{3}{(1+2\bar{\rho}_{0,k}\bar{\lambda}_k)^2}\right)\right]}. \nonumber\\
\eea
The flow of $\bar{\lambda}_k$ is analogous to Eq.~(\ref{Eq:lambdaflow}), but $\eta_k^{(0)}$ is replaced by $\eta_k$:
\bea
\label{Eq:lambdaflow2}
k\partial_k \bar{\lambda}_k = -2\bar{\lambda}_k [1 - \eta_k]&+&\frac{\bar{\lambda}_k^2}{2\pi} \left(1-\frac{\eta_k}{4}\right) \nonumber\\
&\times& \left[1+\frac{9}{(1+2\bar{\rho}_{0,k}\bar{\lambda}_k)^3}\right].
\eea
Now we solve the coupled equations (\ref{Eq:eta0flow}), (\ref{Eq:wkflow}), (\ref{Eq:rhoflow2}) and (\ref{Eq:lambdaflow2}). The corresponding flow diagram can be seen on the right-hand side of Fig.~\ref{Fig:flows}. The comparison shows that taking into account the derivative of the wave function renormalization factor in the anomalous dimension significantly stabilizes the flow along the line of (quasi-)fixed points, as in the improved case the freezing of the flow holds on $\sim 20$ times longer in RG time $t=-\log(k/\Lambda)$.

\begin{figure*}
\begin{center}
\includegraphics[bb = 10 0 255 155,scale=1,angle=0]{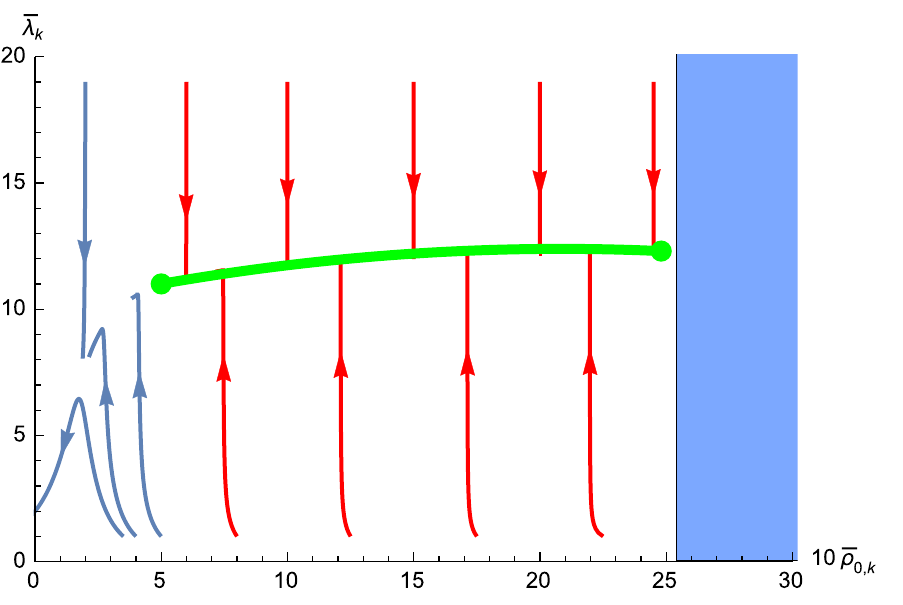}
\caption{Flow diagram for the modified $XY$ model with the initial condition $w_\Lambda=0.4$. The red curves end on the line of fixed points, while the blue ones deviate from it. The fixed line is terminating at two endpoints, the left one corresponding to the high-temperature transition (BKT) and the right one controlling the low-temperature transition. The position of the latter depends on the initial value $w_\Lambda$ (note that the position of the former is not sensitive to $w_\Lambda$, if it exists). The flows become divergent in the shaded region of the parameter space in accordance with (\ref{Eq:ineq}).}
\label{Fig:newflow}
\end{center}
\end{figure*}

\subsection{Phase structure}

Now we are in a position to show that in the modified $XY$ model fluctuations can dramatically change the structure of the line of fixed points, 
as seen in Fig.~\ref{Fig:flows}. First, note that, the ansatz of Eq.~(\ref{Eq:LPAp}) and the approximation $Z_k(\bar{\rho})\approx Z_k(\bar{\rho}_{0,k})+Z_k'(\bar{\rho}_{0,k})(\bar{\rho}-\bar{\rho}_{0,k})$ is compatible with the microscopic Hamiltonian of the modified $XY$ model, since from Eq.~(\ref{Eq:LPAp}) we have
\bea
\label{Eq:LPApp}
\Gamma_k = \int d^2x &&\!\!\!\!\!\!\!\Big[ \frac{Z_k+Z_k^2w_k({\rho}-{\rho}_{0,k})}{2}  (\nabla {\psi}^i)^2 \nonumber\\
&+& \frac{\lambda_k}{2}({\rho}-{\rho}_{0,k})^2\Big],
\eea
which is equivalent to
\bea
\label{Eq:LPApp2}
\Gamma_k = \int d^2x &&\bigg[  a_k(\nabla {\psi}^i)^2+4b_k({\psi}^j)^2(\nabla {\psi}^i)^2\nonumber\\
&&+\frac{\lambda_k}{2}\left(\rho-{\rho}_{0,k}\right)^2\bigg],
\eea
where $a_k=(Z_k-Z_k^2w_k{\rho}_{0,k})/2$ and $b_k=Z_k^2w_k/16$. Eq. ~(\ref{Eq:LPApp2}) is now of the form of the original Hamiltonian in Eq.~(\ref{Eq:Ham}) using the $\psi^i$ vector notation.

The reason why the RG flows of the ordinary $XY$ model can change dramatically is that depending on the initial value $w_\Lambda$ (or $b_\Lambda$, equivalently) at the UV scale, $\bar{\rho}_{0,k}$ can approach a singularity, which sends the flows in the $\bar{\lambda}_k$-$\bar{\rho}_k$ plane away from the line of fixed points. What essentially happens is that the line of quasifixed points terminates also at another end point (Fig.~\ref{Fig:newflow}). The end point on the left corresponds to a BKT transition at higher temperature and the new one on the right signals another transition at lower temperature. Even though the method does not make a definite prediction, this should correspond to the Ising transition already reported in earlier papers \cite{lee85,korshunov85,carpenter89}.

Analyzing the flow of $\bar{\rho}_k$, we note that already in the ordinary $XY$ model, i.e., for $w_\Lambda=0$, at first sight it might seem possible that the denominator on the right-hand side of Eq.~(\ref{Eq:rhoflow2}) becomes zero, but it turns out that this never happens. The flow equation always makes $w_k$ decrease as fluctuations are integrated out, and therefore the flows are regular. Note, however, if at the microscopic scale, $w_\Lambda>0$, then $k\partial_k \bar{\rho}_{0,k}$ can indeed blow up.

The condition that needs to be met for a diverging flow is
\bea
\label{Eq:ineq}
w_\Lambda^{-1}<\bar{\rho}_{0,\Lambda}+\frac{1}{16\pi}\left(1+\frac{3}{(1+2\bar{\rho}_{0,\Lambda}\bar{\lambda}_\Lambda)^2} \right),
\eea
which shows that for positive $w_\Lambda$ values the line of fixed points can also terminate on the right (see Fig.~\ref{Fig:newflow}), leading to a two-step transition. For later reference, just as in Sec.~\ref{Sec:basics}, we restrict ourselves to the case
\bea
\label{Eq:a-b-restriction}
a_\Lambda^2+b_\Lambda^2=1,
\eea
i.e., we may use the parametrization $a_\Lambda=\cos \theta$ and $b_\Lambda=\sin \theta$ $(\theta \in [0,\pi/2])$, which leads to the following constraints:
\bea
\cos \theta &=& Z_\Lambda(1-Z_\Lambda w_\Lambda\rho_{0,\Lambda})/2, \\
\sin \theta &=& Z_\Lambda^2w_\Lambda/16.
\eea
Solving them for $w_\Lambda$ and $Z_\Lambda$, we get
\bea
Z_\Lambda=2(\cos\theta+8\rho_{0,\Lambda}\sin\theta), \\
w_\Lambda=\frac{4\sin\theta}{(\cos\theta+8\rho_{0,\Lambda} \sin\theta)^2}.
\eea
Dropping the last term in the large parantheses of the right-hand side of Eq.~(\ref{Eq:ineq}) (we are interested in a rough estimate), we can get the following condition for the critical value of $\rho_{0,\Lambda}$ belonging to the second endpoint of the line of (quasi)fixed points:
\bea
0&=&-\frac{(\cos\theta+8\rho_{0,\Lambda} \sin\theta)^2}{4\sin\theta} \nonumber\\
&+&2(\cos\theta+8\rho_{0,\Lambda}\sin\theta){\rho}_{0,\Lambda}+\frac{1}{16\pi}.
\eea
For a given $\theta$, we solve this equation for $\rho_{0,\Lambda}$ (see the endpoint on the right-side of Fig.~\ref{Fig:newflow}). Surprisingly, if $\theta\neq 0$ is small, i.e. we are close to the $XY$ model, the solution $\rho_{0,\Lambda}|_{\sol}$ is always negative. This means that since the flows blow up for initial values $\rho_{0,\Lambda}>\rho_{0,\Lambda}|_{\sol}$, unless $\bar{\rho}_{0,\Lambda}|_{\sol} \equiv Z_\Lambda \rho_{0,\Lambda}|_{\sol} \approx 0.5$ (which is the location of the original endpoint of the BKT transition), the line of fixed points completely disappears. The critical angle at which this happens is
\bea
\theta_c \approx 86.8^{\circ}.
\eea
That is to say, for $0\neq\theta<\theta_c$, if there is a transition in the system, it cannot be of topological type, no matter how close we are to the $XY$ model (still, at $\theta=0$ we have one, and only one BKT transition). However, once $\theta>\theta_c$, the line of fixed points starts to return to the picture, now equipped with another end point, which indicates that there exist two transitions. A higher-temperature transition has to be of BKT type and a lower-temperature transition, presumably an Ising transition \cite{carpenter89}, is expected to be of second order. Note that the aforementioned structure heavily relies on the assumption $a_\Lambda^2+b_\Lambda^2=1$. Had we not had this constraint and just set, e.g., $a_\Lambda\equiv 1$, we would have found a two-step transition for $0<b<b_c$ (the higher-temperature one being topological), and no topological transition for $b>b_c$ (here $b_c>0$ is some positive constant).

%%%%%%%%%%%%%%%%%%%%%%%%%%%%%%%%%%%%%
\section{Numerical simulations}\label{sec:numerics}

In this section we numerically investigate the equilibrium properties of the modified Goldstone model defined in Eq.~\eqref{Eq:Ham}.

\subsection{Preparation}
The discretized Hamiltonian ${\cal H}_{\Delta x}$ from Eq.~\eqref{Eq:Ham} becomes
\begin{align}
\label{Eq:Ham-discretized}
\begin{split}
& \mathcal{H}_{\Delta x} = \mathcal{H}_1 + \mathcal{H}_{2} , \\
& \mathcal{H}_1 = a \sum_{\langle i,j \rangle} |\psi_i - \psi_j|^2 + b \sum_{\langle i,j \rangle}|\psi_i^2 - \psi_j^2|^2, \\
& \mathcal{H}_2 = \frac{\lambda \Delta x^2}{2} \sum_i (|\psi_i|^2 / 2 - \rho_0)^2,
\end{split}
\end{align}
where $\psi_i$ is the field $\psi$ at the discretized point $\Vec{x} = \Vec{x}_i$ and $\Delta x$ is the lattice spacing (which serves as an ultraviolet cutoff scale).
In the limit of $\lambda \to \infty$ and rewriting $\psi = \sqrt{2 \rho_0} e^{i \theta_i}$, the discretized Hamiltonian $\mathcal{H}_{\Delta x}$ becomes equivalent to the Hamiltonian $\mathcal{H}_{\rm mXY}$ in Eq.~\eqref{Eq:modXY} for the modified $XY$ model with $J = 4 a \rho_0$ and $J^\prime = 8 b \rho_0^2$.

Now we numerically calculate equilibrium ensemble averages
\begin{align}
\langle f \rangle = \frac{\displaystyle \int \left( \prod_{i} d\psi_i d\psi_i^\ast \right) \sum_i f e^{-\mathcal{H}_{\Delta x} / T}}{\displaystyle \int \left( \prod_{i} d\psi_i d\psi_i^\ast \right) \sum_i e^{-\mathcal{H}_{\Delta x} / T}},
\end{align}
by using the Monte Carlo technique.
First, by fixing the amplitude $|\psi_i|$ of the field, we use the cluster Monte Carlo technique with the Wolff algorithm \cite{Wolff}. Then, to accelerate the equilibration process, we alternately apply the Wolff algorithm to equilibrate the phase $\theta_i = \mathrm{arg}[\psi_i]$ and the standard Metropolis-Hastings algorithm to equilibrate the amplitude $|\psi_i|$.
For numerical parameters, we have used $\Delta x = 1$, $\rho_0 = 1/2$.
Similarly to the preceding section, we parametrize $a$ and $b$ as in 
Eq.~\eqref{Eq:a-b-restriction},
\begin{align}
a = \cos\theta, \quad b = \sin\theta, \quad a^2 + b^2 = 1.
\end{align}

\subsection{Correlation function and transition temperature}

We first show our results for the two correlation functions
\begin{align}
\begin{split}
& G_1(r) = \sum_i \sum_{r \leq |x_j| < r + \Delta x} \frac{\displaystyle \Delta x^2 \langle \psi^\ast_{i+j} \psi_i \rangle}{N(r) L^2}, \\
& G_2(r) = \sum_i \sum_{r \leq |x_j| < r + \Delta x} \frac{\displaystyle \Delta x^2 \langle \psi^{\ast\: 2}_{i+j} \psi_i^2 \rangle}{N(r) L^2},
\end{split}
\end{align}
where $L$ is the system size and $N(r)$ is the number of points, $x_i$, that satisfy $r \leq |x_i| < r + \Delta x$.
When $\theta = 0$ ($\theta = \pi/2$), we expect the standard BKT transition triggered by integer vortices (half-quantized vortices) for $\psi_i$ ($\psi_i^2$) and the algebraic decay $G_1(r) \propto r^{-\eta}$ [$G_2(r) \propto r^{-\eta}$] below the BKT transition temperature.
At the BKT transition temperature, the critical exponent satisfies $\eta = 1/4$ \cite{kosterlitz71,kosterlitz72}.
To obtain the BKT transition temperature, therefore, we can use the finite-size scaling of the correlation functions, in which $G^{(1,2)} / r^{-1/4}$ is expected to be a universal function of $r/L$.
\begin{figure}[htb]
\centering
\includegraphics[height=0.55\linewidth]{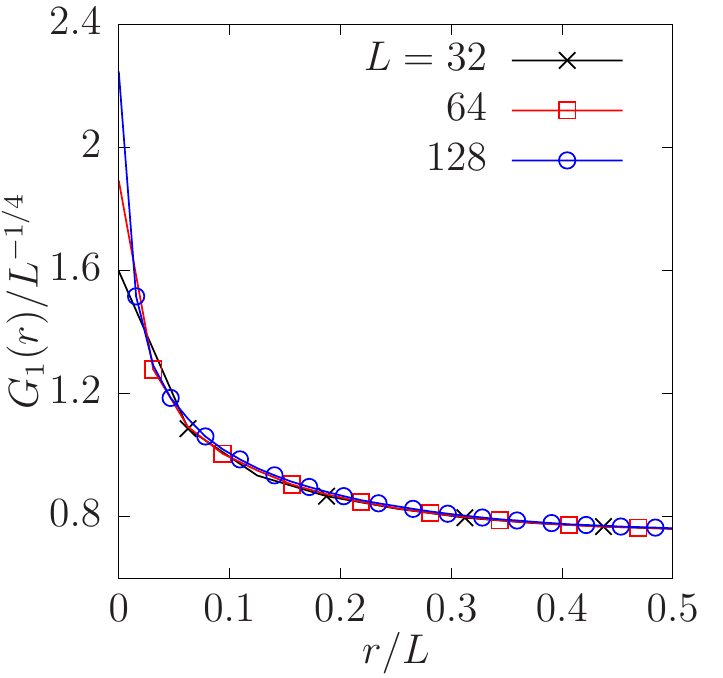}
\caption{
\label{Fig:correlation1}
Finite-size scaling of $G_1(r)$ with $\theta = 0$ and $\lambda = 8$ at the BKT transition temperature $T_1^{\rm BKT} = 0.60 T^\ast$ with the critical exponent $\eta = 1/4$.
The system sizes are $L = 32$ (crosses), $L = 64$ (open squares), and $L = 128$ (open circles).
We use the same symbols for the system size $L$ in all other figures unless otherwise noted.
}
\end{figure}
Figure ~\ref{Fig:correlation1} shows the dependence of $G_1(r) / L^{-1/4}$ with $\theta = 0$ and $\lambda = 8$ as a function of $r / L$ at $T = 0.6 T^\ast$ where $T^\ast$ is the BKT transition temperature for the standard $XY$ model with $\theta = 0$ and $\lambda \to \infty$.
The expected universality of $G_1(r)$ is sufficiently satisfied at large $r$, which, therefore, predicts that the BKT transition temperature is $T_1^{\rm BKT} \simeq  0.6 T^\ast$. In the same way, we can estimate the temperature $T_2^{\rm BKT} \simeq 0.21 T^\ast$ with $\theta = \pi/2$ and $\lambda = 8$ from the finite-size scaling of $G_2(r)$. 

We further expect the appearance of a second-order, Ising-type phase transition \cite{carpenter89}, where the domain of definition for the phase of the $\psi$ field is spontaneously broken from [$0$,$2\pi$] to [$0$,$\pi$], which can be thought of as a spontaneous breaking of a discrete $\mathbb{Z}_2$ symmetry.
At the critical temperature for this phase transition, the correlation function also shows algebraic decay.
Since the critical exponent $\eta$ takes the same value as that of the BKT transition temperature, i.e., $\eta = 1/4$ for the two-dimensional Ising-type transition, we can use the same finite-size scaling analysis as shown in 
Fig.~\ref{Fig:correlation1}.
We here define the temperature $T_1$ ($T_2$) at which $G_1(r)$ [$G_2(r)$] shows the algebraic decay $G_1 \propto r^{-1/4}$ [$G_2(r) \propto r^{-1/4}$].
Then, by definition, $T_1 = T_1^{\rm BKT}$ at $\theta = 0$ and $T_2 = T_2^{\rm BKT}$ at $\theta = \pi/2$.
Denoting by $\theta_1$ and $\theta_2$ critical angles, we have found the following results for $T_{1}$ and $T_{2}$.
\begin{enumerate}
\parskip 0pt
\item
When $\theta$ is small, i.e., $\theta \leq \theta_1$, then $T_1 > T_2$.
\item
When $\theta$ is large, i.e., $\theta_2 < \theta < \pi/2$, then $T_1 < T_2$.
\item
When $\lambda$ is finite, then $\theta_1 < \theta_2$.
For $\theta_1 < \theta \leq \theta_2$, neither $G_1(r)$ nor $G_2(r)$ satisfies $G_{1,2}(r) \propto r^{-1/4}$ at any temperatures and both $T_1$ and $T_2$ are absent.
\item
When $\lambda \to \infty$ for the modified $XY$ model, then $\theta_1 = \theta_2$, i.e., both $T_1$ and $T_2$ always exist at any $\theta$.
\end{enumerate}
\begin{table}[htb]
\begin{tabular}{c|cc}
$\lambda$ & $\theta_1$ & $\theta_2$ \\ \hline
$8$ & $50.8^\circ$ & $84.5^\circ$ \\
$16$ & $66.0^\circ$ & $79.6^\circ$ \\
$\infty$ & \multicolumn{2}{c}{$64.2^\circ$} \\
\end{tabular}
\caption{
\label{TABLE:transition-theta}
Specific values of $\theta_1$ and $\theta_2$ at $\lambda = 8$, $16$, and $\infty$ (modified $XY$ model).
}
\end{table}
The specific values of $\theta_1$ and $\theta_2$ are shown in Table \ref{TABLE:transition-theta}.

\subsection{Superfluid density and specific heat}

To determine the type of the transitions, we calculate the superfluid density $\rho_{\rm s}$ defined as \cite{Chaikin,Thijssen}
\begin{align}
\rho_{\rm s} = \frac{1}{(a + 4 b) L^2} \lim_{\delta \to 0} \frac{F(\delta) - F(0)}{\delta^2}
\end{align}
and the specific heat $C = d\langle \mathcal{H} \rangle / dT$, where $F(\delta) = - T \log \langle e^{- \mathcal{H} / T} \rangle$ is the free energy under the argument-twisted boundary condition $\psi(x + L) = e^{i \delta \cdot L} \psi(x)$.
When a BKT transition occurs at the transition temperature $T^{\rm BKT}$, the universal jump $\Delta \rho_{\rm s}$ of the superfluid density is
\begin{align}
\Delta \rho_{\rm s} = \frac{T^{\rm BKT}}{\pi}.
\label{Eq:universal-relation}
\end{align}
On the other hand, for second-order transitions we expect close to the corresponding critical temperature ($T^{\rm 2nd}$) that the superfluid density obeys $\rho_{\rm s} \propto (T^{\rm 2nd} - T)^\zeta$.
The critical exponent $\zeta$ is obtained by the Josephson relation $\zeta = 2 \beta - \nu \eta$, where $\beta$, $\nu$, and $\eta$ are the critical exponents of the order parameter, the correlation length, and the correlation function, respectively.
By inserting $\beta = 1/8$, $\nu = 1$, and $\eta = 1/4$ for the Ising-type transition, we obtain $\zeta = 0$, i.e., the superfluid density also jumps at the transition temperature, similarly to the BKT transition. However, no universal relation holds, which allows for a distinction between the two.

\begin{figure}[htb]
\centering
\begin{minipage}{0.494\linewidth}
\centering
\includegraphics[height=0.99\linewidth]{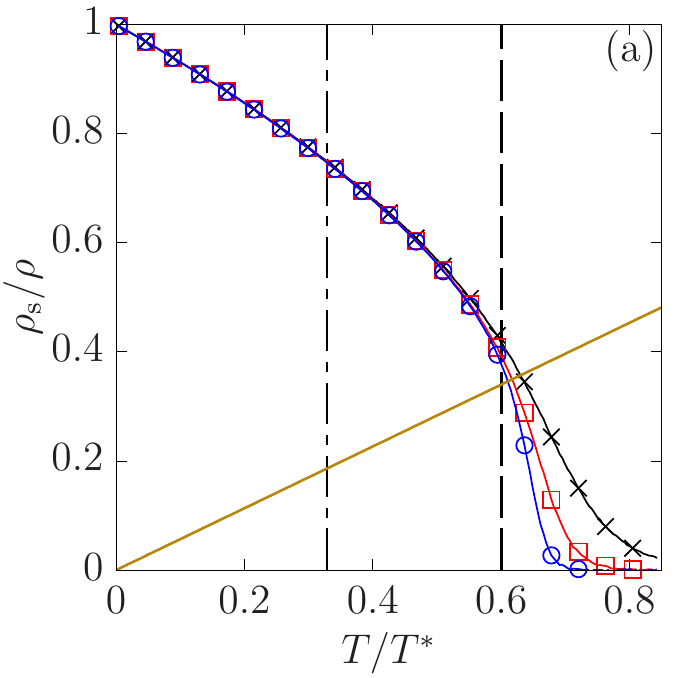}
\end{minipage}
\begin{minipage}{0.494\linewidth}
\centering
\includegraphics[height=0.99\linewidth]{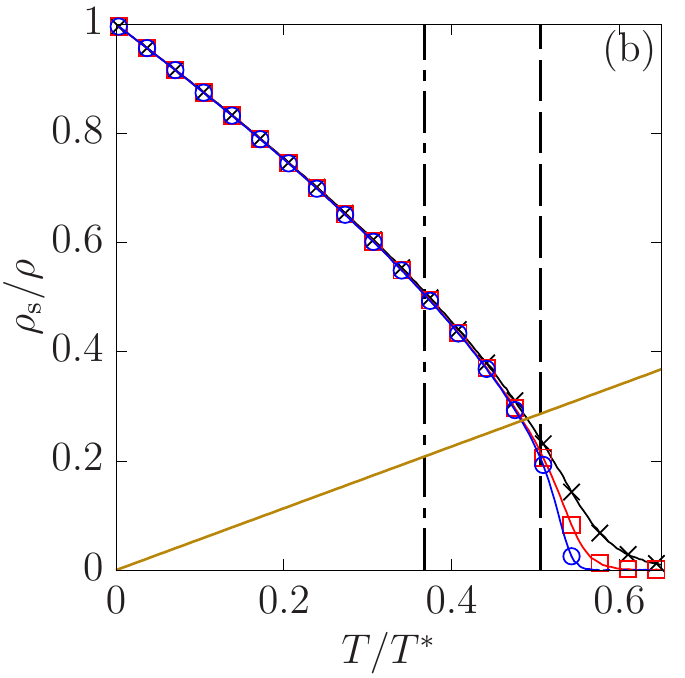}
\end{minipage}
\caption{
\label{Fig:rhos-small-08}
Temperature dependence of the superfluid density $\rho_{\rm s}$ for $\lambda = 8$ at  (a) $\theta = 0^\circ$ and (b) $\theta = 10^\circ$ [the panel (b)].
The solid, dashed, and dash-dotted lines show the relation $\rho_{\rm s} = T / \pi$, where $T = T_1$, and $T = T_2$, respectively.
}
\end{figure}
Figure ~\ref{Fig:rhos-small-08} shows the dependence of the superfluid density with respect to the temperature for $\theta = 0^\circ$ [Fig. 6(a)] and $\theta = 10^\circ$ [Fig. 6(b)].
The solid line shows the relation $\rho_{\rm s} = T / \pi$.
In Fig. 6(a) this line intersects $\rho_{\rm s}$ with a good accuracy at $T_1$, suggesting the standard universal relation related to the BKT transition temperature, i.e., we indeed observe a topological transition.
In Fig. 6(b), however, $\rho_{\rm s}$ deviates from the aforementioned line at $T_1$ and therefore we expect that the transition is of second order, with a nonuniversal jump at the transition temperature. Here we relabel $T_1 \equiv T_1^{\rm 2nd}$.
In neither of the panels do we find any characteristic structure in $\rho_{\rm s}$ at $T = T_2$.
We therefore conclude that the property of the correlation function $G_2 \propto r^{-1/4}$ is just the crossover and we relabel $T_2$ as the crossover temperature $T_2 \equiv T_2^{\rm CO}$.
\begin{figure}[htb]
\centering
\begin{minipage}{0.494\linewidth}
\centering
\includegraphics[height=0.99\linewidth]{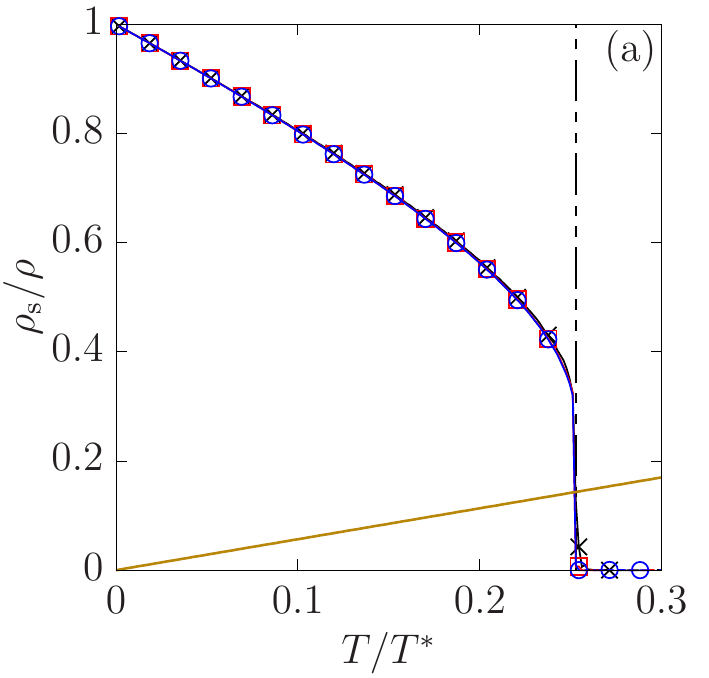}
\end{minipage}
\begin{minipage}{0.494\linewidth}
\centering
\includegraphics[height=0.99\linewidth]{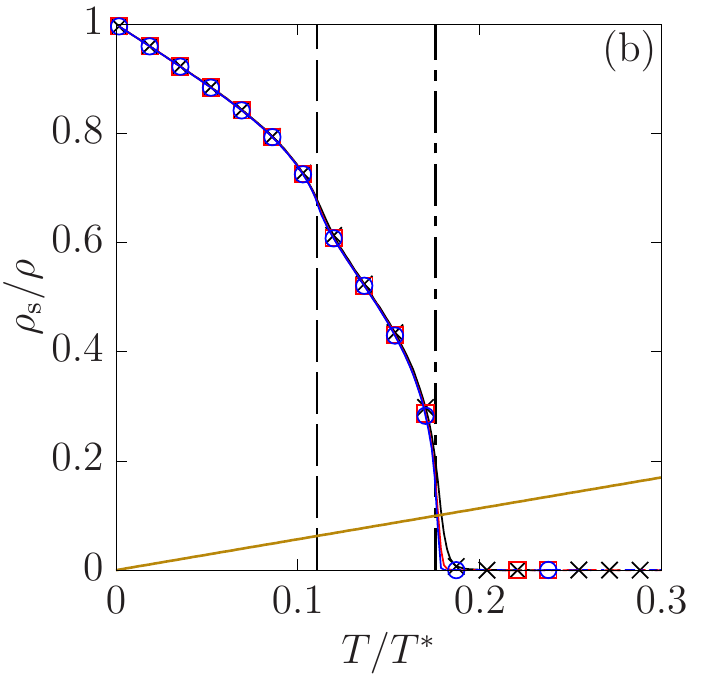}
\end{minipage}
\caption{
\label{Fig:rhos-large-08}
Temperature dependence of the superfluid density $\rho_{\rm s}$ for $\lambda = 8$ and (a) $\theta = 60^\circ$ and (b) $\theta = 87^\circ$.
The solid line shows the relation $\rho_{\rm s} = T / \pi$.
The dash-double-dotted line in (a) shows the estimated first-order transition temperature $T_\ast^{\rm 1st}$.
The dashed and the dash-dotted lines (b) show $T = T_1$ and $T = T_2$, respectively.
}
\end{figure}

Figure ~\ref{Fig:rhos-large-08} shows the dependence of the superfluid density on the temperature for $\theta = 60^\circ$ [Fig. 7(a)] and $\theta = 85^\circ$ [Fig. 7(b)].
As shown in Table \ref{TABLE:transition-theta}, the value $\theta = 60^\circ$ is between $\theta_1$ and $\theta_2$ for $\lambda = 8$, and we find neither a BKT nor a second-order phase transition.
Instead, what we see is a first order phase transition due to the sharp jump of the superfluid density $\rho_{\rm s}$ [see Fig.~\ref{Fig:rhos-large-08} (a)].
Because the temperature at which the superfluid density $\rho_{\rm s}$ jumps does not really depend on the system size $L$, its estimation is fairly simple. We denote this transition temperature by $T_\ast^{\rm 1st}$.
In Fig.~\ref{Fig:rhos-large-08} (b), i.e., for $\theta = 87^\circ$, $\theta$ is larger than $\theta_2$ and the superfluid density $\rho_{\rm s}$ does show the universal relation \eqref{Eq:universal-relation} at the corresponding temperature, $T = T_2$. Therefore, we find again a BKT transition with the aforementioned transition temperature, relabeling it as $T_2 \equiv T_2^{\rm BKT}$.

\begin{figure}[htb]
\centering
\begin{minipage}{0.494\linewidth}
\centering
\includegraphics[height=0.95\linewidth]{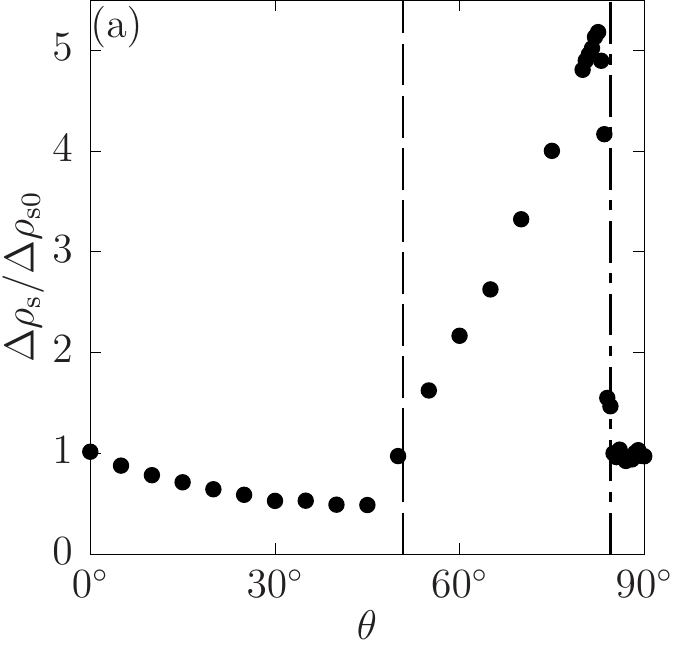}
\end{minipage}
\begin{minipage}{0.494\linewidth}
\centering
\includegraphics[height=0.95\linewidth]{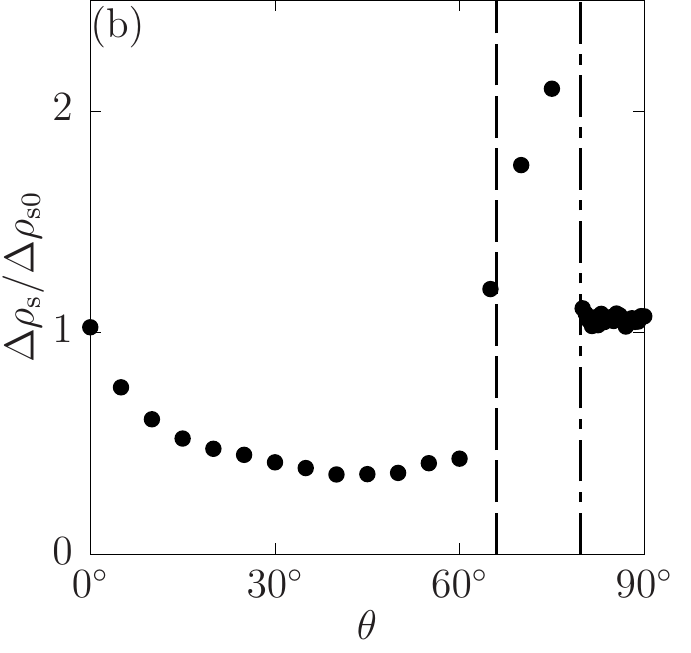}
\end{minipage}
\caption{
\label{Fig:rhoc}
Jump of the superfluid density $\Delta \rho_{\rm s}$ normalized by $\Delta \rho_{\rm s0}$ for (a) $\lambda = 8$ and (b) $\lambda = 16$.
The dashed and dash-doted lines show $\theta_1$ and $\theta_2$, respectively.
}
\end{figure}
Figure ~\ref{Fig:rhoc} shows the jump of the superfluid density $\Delta \rho_{\rm s}$ at the phase transition as a function of $\theta$, normalized by $\Delta \rho_{\rm s0}$, which is the value for the universal jump \eqref{Eq:universal-relation} for the BKT transition. It is specifically defined as (note that $T_1$, $T_*^{\rm 1st}$, and $T_2$ depend on $\theta$)
\begin{align}
\Delta \rho_{\rm s0} = \begin{cases}
\displaystyle \frac{T_1}{\pi}, & \displaystyle 0 \leq \theta < \theta_1 \\[5pt]
\displaystyle \frac{T_\ast^{\rm 1st}}{\pi}, & \displaystyle \theta_1 \leq \theta < \theta_2 \\[5pt]
\displaystyle \frac{T_2}{\pi}, & \displaystyle \theta_2 \leq \theta \leq \frac{\pi}{2}.
\end{cases}
\end{align}
We estimate the value of the jump $\Delta \rho_{\rm s}$ by fitting the superfluid density $\rho_{\rm s}$ at the transition temperature, i.e. $T_1$ for $0 \leq \theta < \theta_1$, $T_\ast^{\rm 1st}$ for $\theta_1 \leq \theta < \theta_2$, and $T_2$ for $\theta_2 \leq \theta \leq \pi/2$, via the function
\begin{align}
\rho_{\rm s}(\theta,L) = \Delta \rho_{\rm s}(\theta) + \frac{a(\theta)}{L},
\end{align}
where $a$ is a $\theta$-dependent constant.
For $\theta = 0$ and $\theta > \theta_2$, the relation $\Delta \rho_{\rm s} \simeq \Delta \rho_{\rm s0}$ is satisfied; therefore, we find BKT transitions with the transition temperature $T_1^{\rm BKT}$ for $\theta = 0$ and $T_2^{\rm BKT}$ for $\theta_1 \leq \theta \leq \pi/2$.
For other values, the universal relation does not hold and the transition becomes of second order for $0 < \theta < \theta_1$, and of first order for $\theta_1 < \theta \leq \theta_2$.

\begin{figure}[htb]
\centering
\begin{minipage}{0.494\linewidth}
\centering
\includegraphics[height=0.99\linewidth]{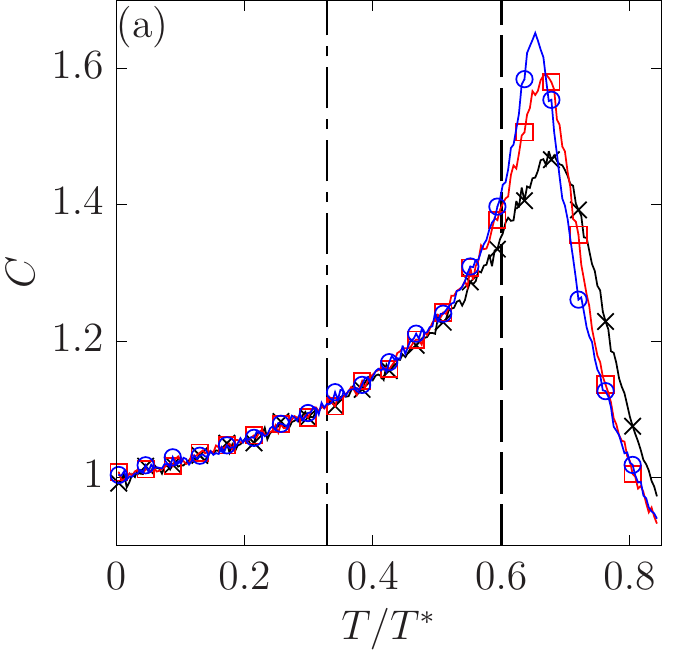}
\end{minipage}
\begin{minipage}{0.494\linewidth}
\centering
\includegraphics[height=0.99\linewidth]{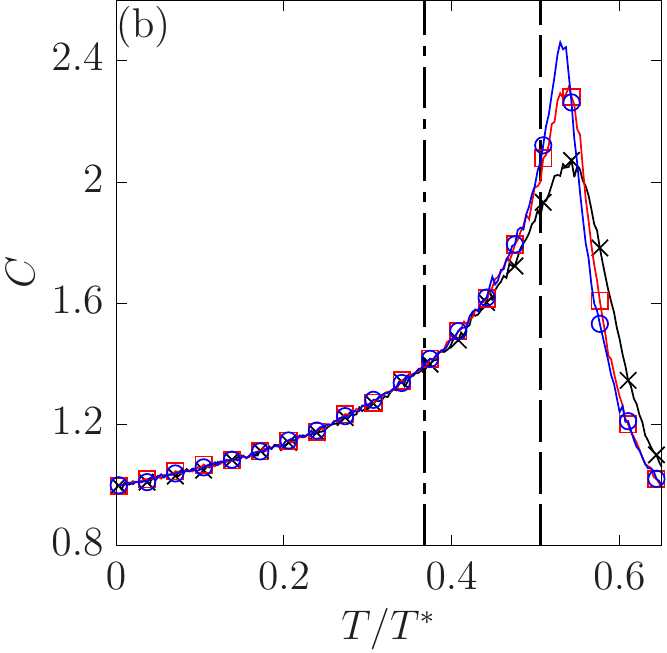}
\end{minipage} \\[5pt]
\begin{minipage}{0.494\linewidth}
\centering
\includegraphics[height=0.99\linewidth]{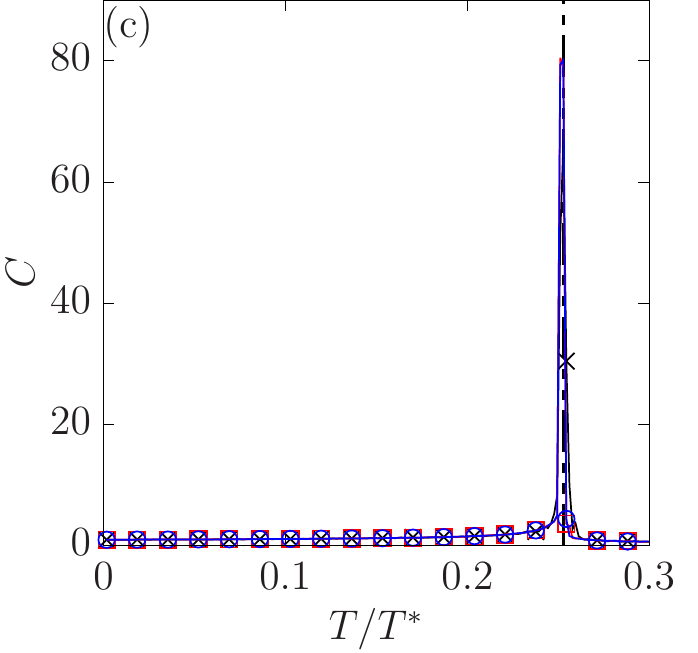}
\end{minipage}
\begin{minipage}{0.494\linewidth}
\centering
\includegraphics[height=0.99\linewidth]{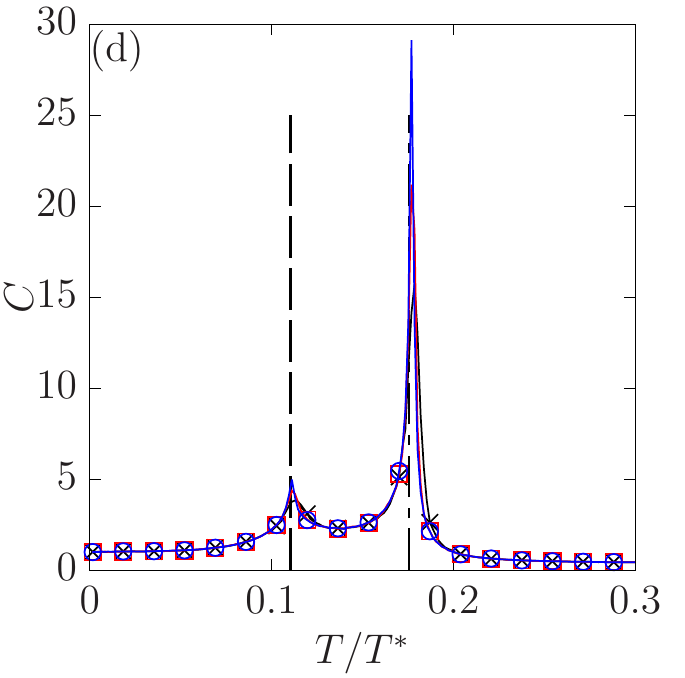}
\end{minipage}
\caption{
\label{Fig:specific-heat-08}
Temperature dependence of the specific heat $C$ for $\lambda = 8$ and (a) $\theta = 0^\circ$, (b) $\theta = 10^\circ$, (c) $\theta = 60^\circ$, and (d) $\theta = 87^\circ$.
The dash-double-dotted line in (c) shows the estimated first-order transition temperature $T_\ast^{\rm 1st}$.
The dashed and dash-dotted lines in (a), (b), and (d) show $T = T_1$ and $T = T_2$, respectively.
}
\end{figure}

Figure ~\ref{Fig:specific-heat-08} shows the specific heat $C$.
Whereas the specific heat has a single peak near the transition temperature for $\theta < \theta_2$, i.e., in Figs. 9(a)-(c), it has double peaks for $\theta \geq \theta_2$, suggesting two-step transitions.
In the latter case, the first and second peaks of the specific heat correspond to the temperatures $T_1$ and $T_2$, respectively.
Because the correlation function $G_1$ becomes $G_1 \propto r^{-1/4}$ at $T = T_1$ and the phase at $T < T_1$ should be continuously connected from the phase with $\theta < \theta_2$ (see Fig.~\ref{Fig:phase-diagram}), the transition at $T_1$ should indeed be of second order.
The absence of the peak at $T = T_2$ for $\theta < \theta_1$ consolidates our conclusion that here $T_2$ gives not the transition, but only a crossover as $T_2^{\rm CO}$.

%%%%%%%%%%%%%
\subsection{Phase diagram}

\begin{figure}[htb]
\centering
\begin{minipage}{0.494\linewidth}
\centering
\includegraphics[height=0.99\linewidth]{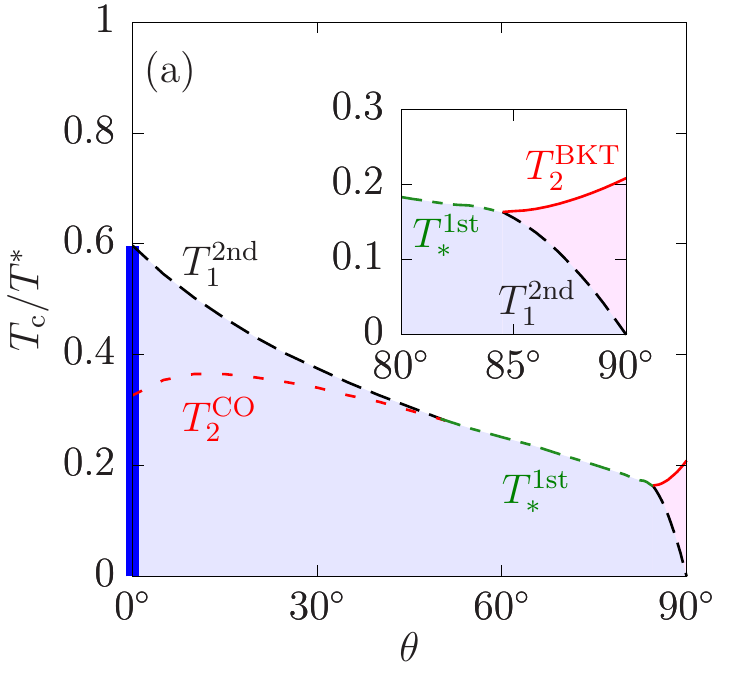}
\end{minipage}
\begin{minipage}{0.494\linewidth}
\centering
\includegraphics[height=0.99\linewidth]{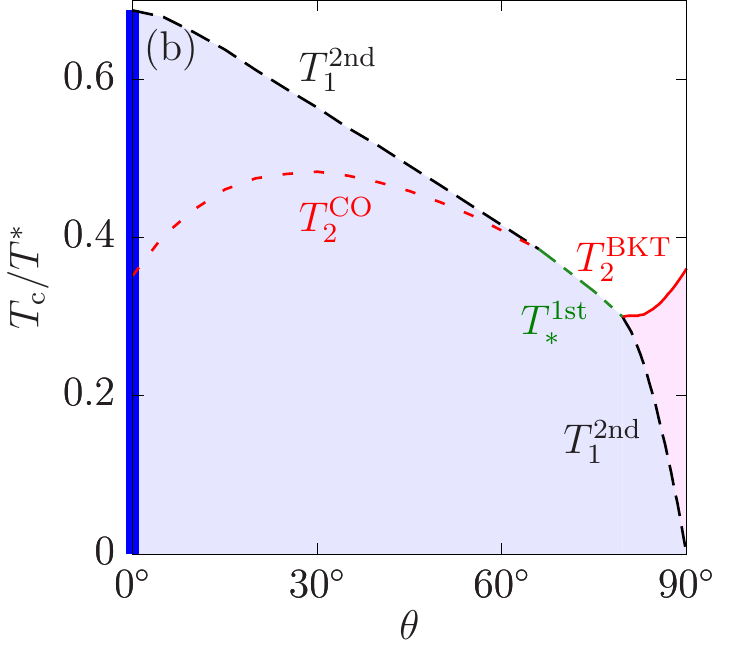}
\end{minipage}
\caption{
\label{Fig:phase-diagram}
Phase diagram in the $\theta$-$T$ plane for (a) $\lambda = 8$ and (b) $\lambda = 16$.
The thick line at $\theta = 0$ is the quasi-long-range order phase with bounded integer vortex pairs.
The violet and pink regions indicate the true long-range-order phase and the quasi-long-range order phase with bounded half-quantized vortex pairs, respectively.
The solid, dashed, and dash-dotted lines correspond to the phase boundaries for the BKT, second-order, and first-order transition temperatures $T_2^{\rm BKT}$, $T_1^{\rm 2nd}$, and $T_\ast^{\rm 1st}$, respectively.
The dotted line indicates the crossover temperature $T_2^{\rm CO}$.
}
\end{figure}
Figure~\ref{Fig:phase-diagram} shows the phase diagram of the modified Goldstone model in Eq.~\eqref{Eq:Ham-discretized}.
For $\theta = 0$, there is the standard BKT transition with the transition temperature $T_1\equiv T_1^{\rm BKT}$.
At $T < T_1^{\rm BKT}$, integer vortex pairs are bounded to show a quasi-long-range ordered phase.
For $0 < \theta < \theta_1$, this BKT transition changes to a second-order phase transition with the transition temperature $T_1 \equiv T_1^{\rm 2nd}$,
implying a true long-range-order phase for $T<T_1^{\rm 2nd}$ with the breaking of the $\mathbb{Z}_2$ symmetry.
For $\theta_1 \leq \theta < \theta_2$, the two temperatures $T_1^{\rm 2nd}$ and $T_2^{\rm CO}$ defined for $0 < \theta < \theta_1$ merge to one first-order transition temperature, $T_\ast^{\rm 1st}$.
For $\theta_2 \leq \theta \leq \pi/2$, this transition temperature, $T_\ast^{\rm 1st}$, splits again into two transition temperatures $T_1^{\rm 2nd}$ and $T_2^{\rm BKT}$.
The second-order phase transition ultimately disappears, as while $\theta \rightarrow \pi/2$, $T_1^{\rm 2nd} \rightarrow 0$.
Unlike the BKT transition for $\theta = 0$, the BKT transition for $\theta_2 \leq \theta \leq \pi/2$ is triggered by the correlation function $G_2$ (not $G_1$), and therefore we expect the quasi-long-range order phase by the bounding of half-quantized vortex pairs at $T_1^{\rm 2nd} < T < T_2^{\rm BKT}$.
Because the low-temperature phases, i.e., $T < T_\ast^{\rm 1st}$ for $\theta_1 \leq \theta < \theta_2$ and $T < T_1^{\rm 2nd}$ for $\theta_2 \leq \theta \leq \pi/2$, should be continuously connected to the long-range-order phase at $0 < \theta < \theta_1$, these phases should also be of true long-range order.

Here we wish to establish the relationship between the phase diagram and the (quasi-)breaking patterns of symmetry summarized in Eqs.~\eqref{eq:ssb-simple-BKT}-\eqref{eq:ssb-BKT-TT}.
The BKT transition at the temperature $T_1^{\rm BKT}$ with $\theta = 0^\circ$ gives the quasibreaking $\mathrm{U}(1) \dashrightarrow 1$ in Eq.~\eqref{eq:ssb-simple-BKT}.
The second and first-order phase transitions at the temperatures $T_1^{\rm 2nd}$ and $T_\ast^{\rm 1st}$ with $0^\circ < \theta \leq \theta_2$, respectively, give the simultaneous (quasi)breaking $\mathrm{U}(1) \Longrightarrow 1$ in Eq.~\eqref{eq:ssb-BKT-TT}.
The two-step transition at the temperatures $T_2^{\rm BKT}$ and $T_1^{\rm 2nd}$ with $\theta_2 < \theta < 90^\circ$ gives the two successive (quasi)breaking of symmetries $\mathrm{U}(1) \dashrightarrow \mathbb{Z}_2 \longrightarrow 1$ in Eq.~\eqref{eq:ssb-two-step}.
As for $\theta = 90^\circ$, the BKT transition at the temperature $T_2^{\rm BKT}$ gives the quasi-breaking $\mathrm{U}(1)/ \mathbb{Z}_2 \dashrightarrow 1$.
Here the second-order phase transition does not occur because of $T_1^{\rm 2nd} = 0$ for $\theta = 90^\circ$.

\begin{figure}[htb]
\centering
\begin{minipage}{0.494\linewidth}
\centering
\includegraphics[height=0.99\linewidth]{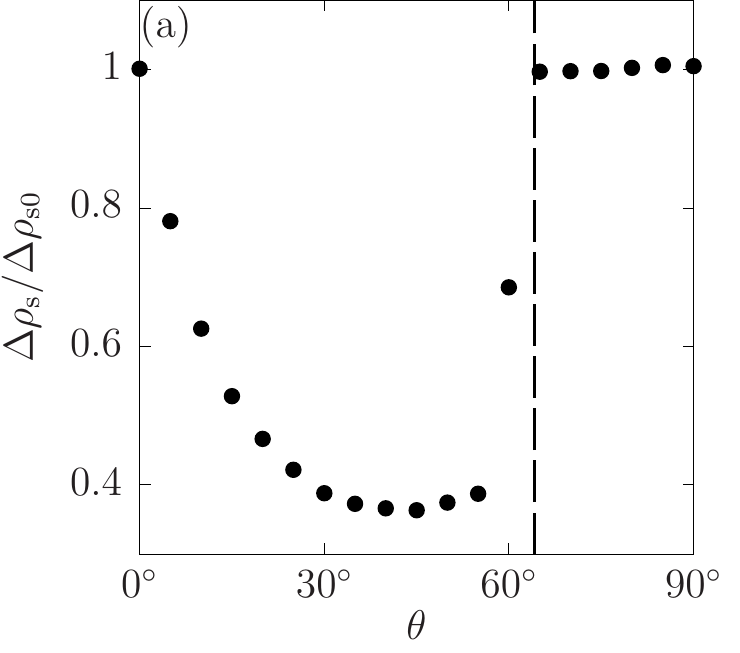}
\end{minipage}
\begin{minipage}{0.494\linewidth}
\centering
\includegraphics[height=0.99\linewidth]{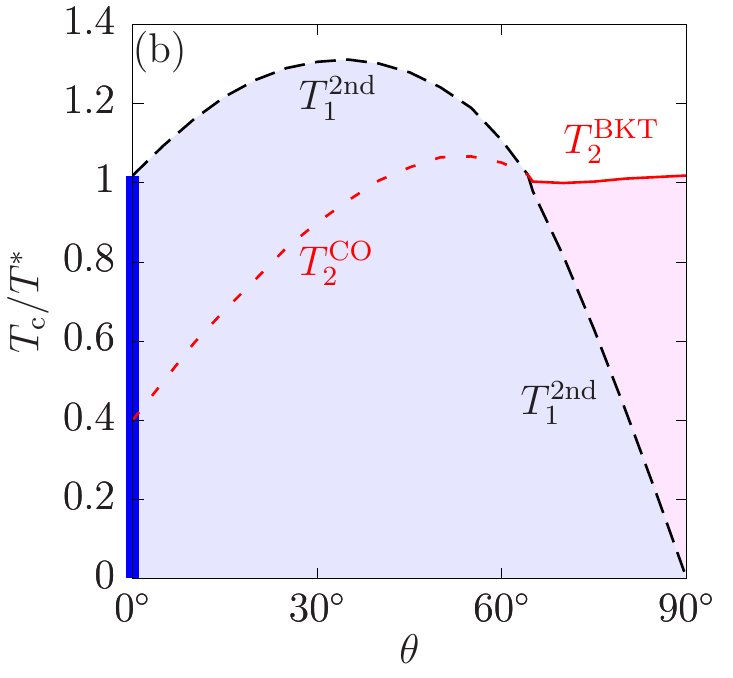}
\end{minipage}
\caption{
\label{Fig:phase-diagram-inf}
(a) Jump of the superfluid density $\Delta \rho_{\rm s}$ and (b) phase diagram in the $\theta$-$T$ plane for $\lambda = \infty$.
The dashed line in (a) and those in the colored regions in (b) are the same as those in Figs.~\ref{Fig:rhoc} and ~\ref{Fig:phase-diagram}, respectively.
}
\end{figure}
Finally, in Fig.~\ref{Fig:phase-diagram-inf}
we show the jump of the superfluid density $\Delta \rho_{\rm s}$ and the phase diagram in the $\lambda = \infty$ limit,
in which the modified Goldstone model reduces to the modified $XY$ model. 
As the coupling $\lambda$ increases, the region of the first-order phase transition for $\theta_1 < \theta \leq \theta_2$ shrinks and ultimately disappears.

%%%%%%%%%%%%
\subsection{Vortex configurations}

Here we discuss the relationship between topological defects (such as integer and half-integer vortices and one-dimensional solitons considered 
in Sec.~\ref{Sec:basics}) and the corresponding phase transitions.
At the BKT transition temperature $T_1^{\rm BKT}$ with $\theta = 0^\circ$, the number of integer vortex-antivortex pairs is changing rapidly due to their bounding.
At the second and first-order transition temperatures $T_2^{\rm 2nd}$ and $T_1^{\rm 1st}$, the $\mathbb{Z}_2$ symmetry breaking causes the rapid decrease of one-dimensional solitons.
At the BKT transition temperature $T_2^{\rm BKT}$ with $\theta > \theta_2$, the number of half-integer vortex-antivortex pairs changes rapidly.
The vortexmolecules, which contain two half-quantized vortices should be stable in order for the BKT transition to exist at the temperature $T_2^{\rm BKT}$.
On the other hand, the stability of one-dimensional solitons is enough for the existence of the $\mathbb{Z}_2$ symmetry breaking.
The stability of vortex molecules for $\theta \gtrsim 78^\circ$ and of one-dimensional solitons for $\theta \gtrsim 15^\circ$ in the case of $\lambda = 8$ is consistent with the existence of $T_2^{\rm BKT}$ for $\theta > \theta_2 \approx 84.5^\circ$, and the $\mathbb{Z}_2$ symmetry breaking at $T_1^{\rm 2nd}$ or $T_\ast^{\rm 1st}$ for $\theta > 0$.

\begin{figure}[htb]
\centering
\begin{minipage}{0.494\linewidth}
\centering
\includegraphics[height=0.99\linewidth]{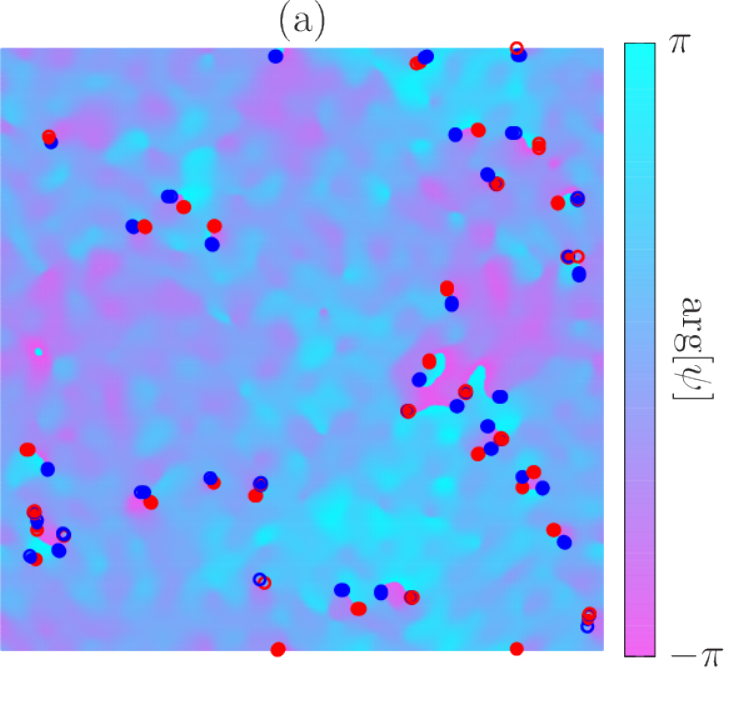}
\end{minipage}
\begin{minipage}{0.494\linewidth}
\centering
\includegraphics[height=0.99\linewidth]{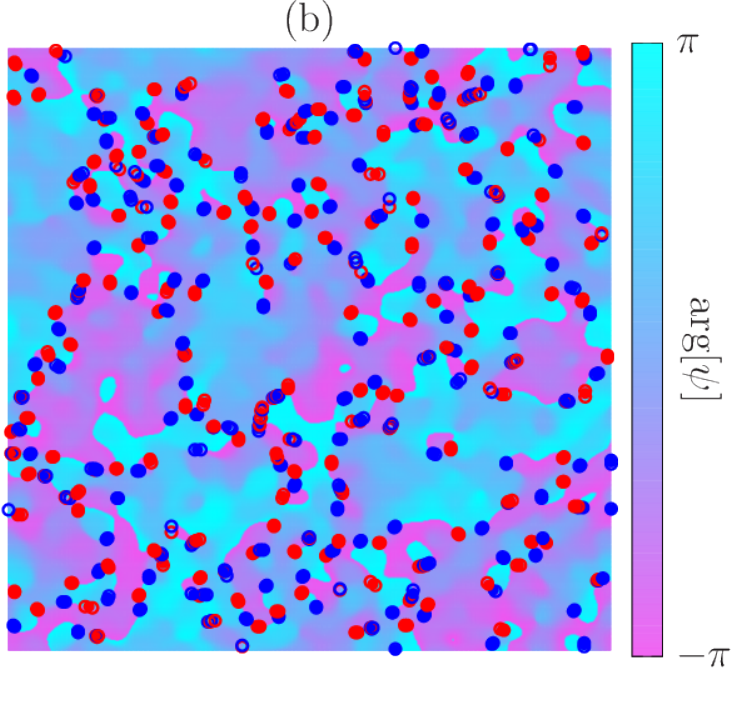}
\end{minipage} \\[10pt]
\begin{minipage}{0.494\linewidth}
\centering
\includegraphics[height=0.99\linewidth]{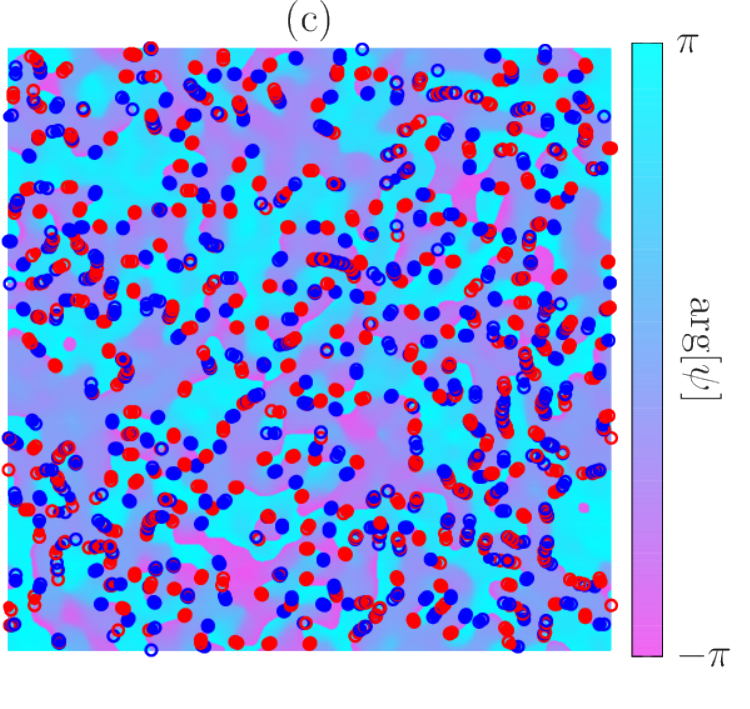}
\end{minipage}
\begin{minipage}{0.494\linewidth}
\centering
\includegraphics[height=0.99\linewidth]{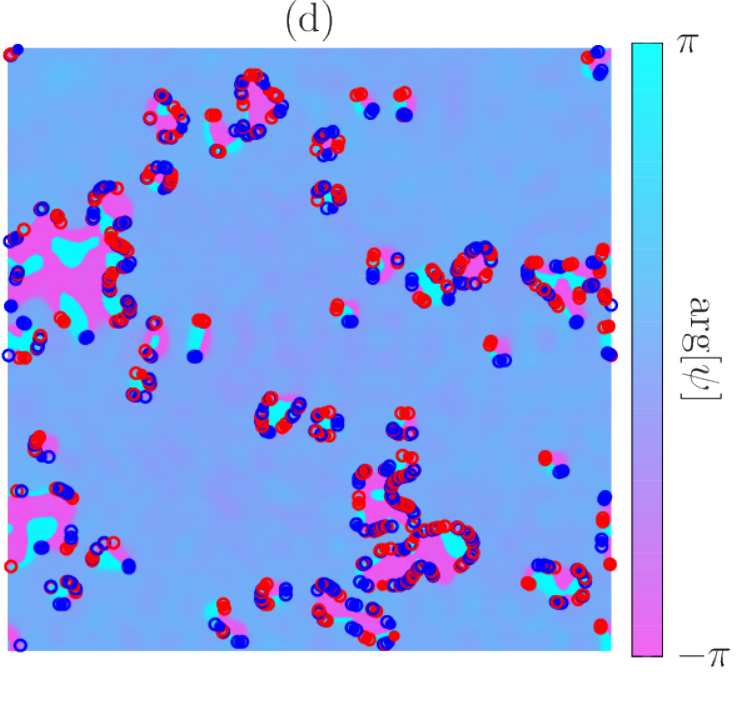}
\end{minipage}
\caption{
\label{Fig:vortex-08}
Snapshots of the vortex configurations and the phase profiles for $L = 64$, $\lambda = 8$ and (a) $\theta = 10^\circ$ and $T = T_1^{\rm 2nd}$, (b) $\theta = 60^\circ$ and $T = T_\ast^{\rm 1st}$, (c) $\theta = 87^\circ$ and $T = T_2^{\rm BKT}$, and (d) $\theta = 87^\circ$ and $T = T_1^{\rm 2nd}$.
The blue and red closed (open) circles denote the positions of integer (half-integer) vortices and antivortices, respectively.
}
\end{figure}
We next show snapshots of vortex configurations and the phase profile at the transition temperatures in Fig.~\ref{Fig:vortex-08}.
In all the panels, most vortices and antivortices form paired states with short distances.
Furthermore, most of them lie on the solitons that appear as boundaries between the two phases $\mathrm{arg}[\psi] \sim 0$ and $\mathrm{arg}[\psi] \sim \pi$.

\begin{figure}[htb]
\centering
\begin{minipage}{0.494\linewidth}
\centering
\includegraphics[height=0.9\linewidth]{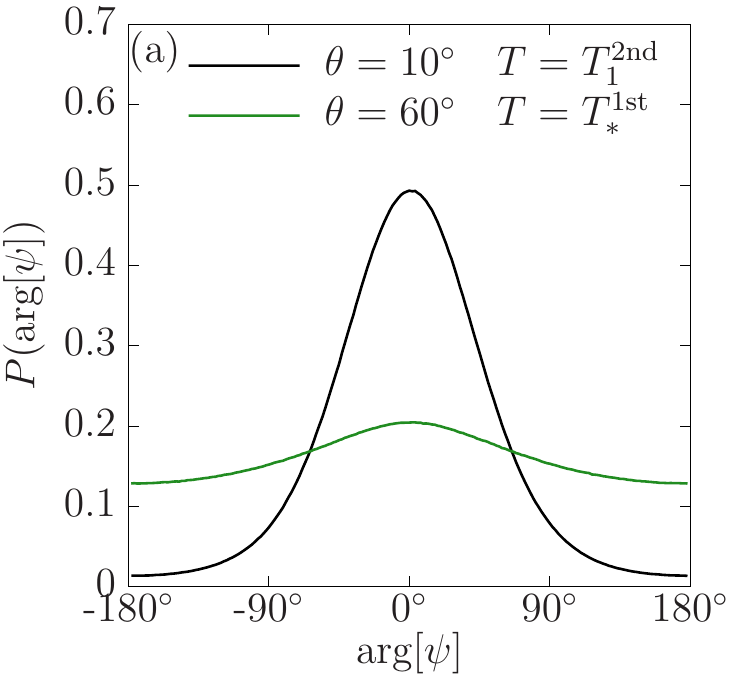}
\end{minipage}
\begin{minipage}{0.494\linewidth}
\centering
\includegraphics[height=0.9\linewidth]{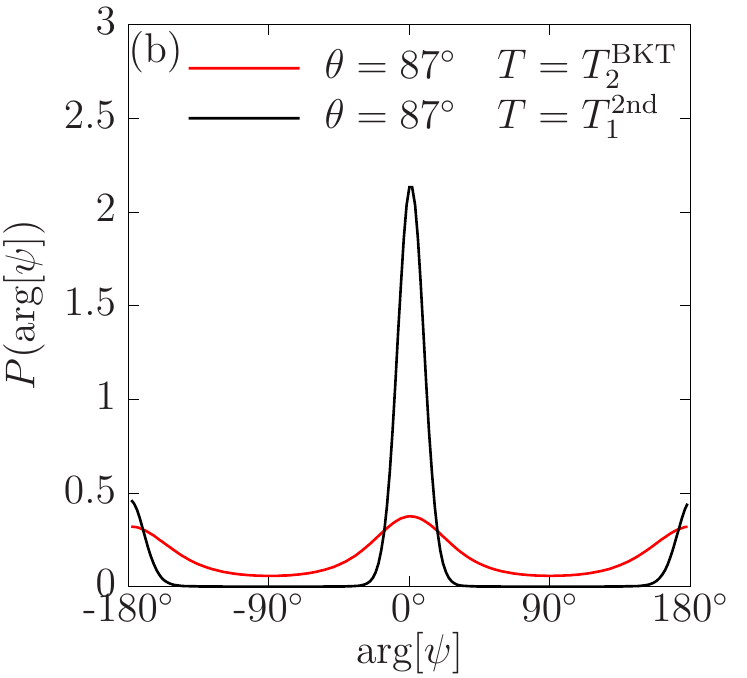}
\end{minipage}
\caption{
\label{Fig:histgram}
Distribution functions $P(\mathrm{arg}[\psi])$ corresponding to the snapshots of the phase profile with $L = 128$, $\lambda = 8$ and (a) $\theta = 10^\circ$ and $T = T_1^{\rm 2nd}$ (black) and $\theta = 60^\circ$ and $T = T_\ast^{\rm 1st}$ (green), and (b) $\theta = 87^\circ$ and $T = T_2^{\rm BKT}$ (red) and (d) $\theta = 87^\circ$ and $T = T_1^{\rm 2nd}$ (black).
}
\end{figure}

Figure ~\ref{Fig:histgram} shows the distribution function $P(\mathrm{arg}[\psi])$ corresponding to the snapshot of the phase profile.
In Fig.~\ref{Fig:histgram} (b) for $\theta = 87^\circ$, the stability of one-dimensional solitons can be clearly seen from the double-peaked structure of $P(\mathrm{arg}[\psi])$ at $\mathrm{Arg}[\psi] = 0$ and $\mathrm{arg}[\psi] = \pi$.
At $T = T_2^{\rm BKT}$, the $\mathbb{Z}_2$ symmetry is not broken and the heights of two peaks are the same.
On the other hand, breaking of the $\mathbb{Z}_2$ symmetry at $T = T_1^{\rm 2nd}$ can be confirmed via the existence of imbalanced peaks $P(0) > P(\pi)$.
This imbalanced distribution can also be seen in Fig.~\ref{Fig:vortex-08} (d), where the region with $\mathrm{arg}[\psi] \sim 0$ is apparently larger than that with $\mathrm{arg}[\psi] = \pi$ and $\mathrm{Arg}[\psi] = -\pi$.
Note that, in Fig.~\ref{Fig:histgram} (a) for $\theta = 10^\circ$ and $\theta = 60^\circ$, however, the double peaked structure is absent and there is only one single peak at $\mathrm{arg}[\psi] = 0$. We believe that this absence comes from finite-size effects and it is expected that the double-peaked structure is restored with larger system size.
We note that all the peaked structures shown in Figs.~\ref{Fig:histgram}(a) and \ref{Fig:histgram}(b) come from finite-size effects and they become completely flat in the thermodynamic limit due to the CMW theorem.

%%%%%%%%%%%%%%%%%%%%%%%%%%%%%%%%%
\section{Summary}\label{sec:summary}

In this paper, 
we first defined the modified Goldstone model in Eq.~(\ref{Eq:Ham}) as a regular and continuum version of the modified $XY$ model and constructed a soliton, an integer vortex and a molecule of half-quantized vortices connected by a soliton.
Then we analyzed the phase structure of the modified
Goldstone model in two dimensions via two different
approaches. First, by using the functional renormalization group technique, we
showed how to describe BKT transitions by
calculating the scale evolution of the effective Hamiltonian. Based on
earlier works, we constructed an approximation scheme of the
RG flow equations, where the field dependence of the
wave function renormalization is taken into account  at the lowest order. In the standard
Goldstone model it has led to a more accurate description of the underlying
structure of a line of fixed points, and it has also turned out to be of
particular importance when one is interested in the role of the modified
kinetic term $\sim |\vec{\nabla} \psi^2|^2$, by revealing a second endpoint of
the line of fixed points. The FRG method predicts that in the modified
model there can exist a two-step phase transition, depending on the ratio
between the coefficients of the standard and modified kinetic terms. It has
also been shown that even if the coefficient of the modified kinetic term is
not large enough to split the phase transition into two, it is capable of
completely destroying its topological nature.

In addition,
this scenario has been verified to great accuracy via full numerical simulation
of the system by the Monte Carlo method. Through predicting critical
temperatures and calculating the superfluid density with the specific heat
numerically, we have confirmed the following properties of the phase
structure. If only the standard or modified kinetic terms are present, the
system undergoes one, and only one phase transition, which is of BKT type,
corresponding to vortex and half-vortex unbinding, respectively. If both
terms are present, depending on the ratio between their coefficients, and
by assuming that their square sum equals unity ($a^2+b^2=1$), there exist
either one or two transitions. If there is only one transition, it is never
topological and can be of both first and second order. If there are two
transitions, then the one corresponding to the higher temperature is of BKT
type, presumably related to half-vortex unbinding, while the other
transition is of Ising type.

It would be interesting to improve upon the present renormalization group
approximation scheme. Since higher-field derivatives of the wave function
renormalization factor could also play an important role for BKT-like
transitions, it would be interesting to derive a tower of equations for the
aforementioned factors, and solve them simultaneously \cite{tetradis94,rose17}. Furthermore, the
present scheme has only predicted the existence of a different end point of the
line of fixed points, which indicated a second transition, but
due to the singular nature of the flows below temperatures of the aforementioned
transition, details of the transition could not have been explored.
It would be particularly important to find a scheme
which can overcome this shortcoming.

The results of this paper can be contrasted to another model admitting 
 a vortex molecule solution of half-quantized vortices connected by a soliton, 
 that is, coherently coupled Bose-Einstein condensates or two-gap superconductors \cite{Kobayashi:2018ezm} and spin-1 spinor Bose-Einstein condensates under the quadratic Zeeman field \cite{Kobayashi:2019}. 
 In this case, a two-step phase transition does not occur 
 when two components are coupled by a Josephson interaction or a quadratic Zeeman field,
 while it can occur when they are decoupled. 
Essential differences between this case and that of
the modified Goldstone model discussed in this paper are yet to be clarified.

As an important application of the modified Goldstone model, we suggest a two-dimensional crystal system, where the perfect crystal is forbidden by the CMW theorem.
In this system, there are two different kinds of topological excitations; a dislocation and a disclination corresponding to spontaneous breaking of the translational and rotational symmetries, respectively.
The dislocation can be topologically equivalent to a pair of disclinations.
The Kosterlitz-Thouless-Nelson-Halperin-Young theory predicts a two-step transition \cite{Halperin,Young}, i.e., the BKT-like transition from the disordered phase to the isotropic hexatic phase in which only the quasi-long-range rotational order appears and disclinations are not bounded into dislocations and the first-order transition from the hexatic phase to the crystal phase having the quasi-long-range rotational order and the long-range translational order.
The mechanism for the two-step transition in this system has almost the same scenario as that for the modified Goldstone model, i.e., dislocations and disclinations correspond to integer and half-integer vortices, respectively.
Whereas a two-step transition has been observed in a two-dimensional colloidal crystal \cite{Zahn}, single first-order transitions have been observed in several two-dimensional crystal systems \cite{Naumovets}, and both single and two-step transitions have been reported in a helium film at low temperatures \cite{Nakamura}, depending on the density of helium atoms.
Our modified Goldstone model \eqref{Eq:Ham} can become a toy model for these systems, i.e., $a/b$ and $\lambda$ correspond to the ratio between the energies of disclinations and dislocations, and the compressibility of the system, respectively, and may give some intuitive guiding principle about the type of phase transitions in this system.

Our study of the modified Goldstone model in two Euclidean dimensions has revealed that there exist two-step phase transitions related to half-quantized vortex molecules connected by domain walls.
It is an open question whether there is any higher dimensional model 
allowing a two-step phase transition.
For instance, in three dimensions, a pair of a monopole and an anti-monopole connected by a string may play a crucial role.

\section*{Acknowledgements}

This work was supported by the Ministry of Education, Culture, Sports, Science (MEXT)-Supported Program for the Strategic Research Foundation at Private Universities ``Topological Science'' (Grant No. S1511006). 
This work was also supported in part by JSPS KAKENHI Grants No.
16H03984 (M.\ K. and M.\ N.), 
19K14713 (C.\ C.), 
 and 18H01217 (M.\ N.).
G.F. was also supported by the Hungarian National Research, Development and Innovation Office (Project No. 127982), by the János Bolyai Research Scholarship of the Hungarian Academy of Sciences, and by the ÚNKP-19-4 New National Excellence Program of the Ministry for Innovation and Technology of Hungary. 
The work of M.\ N. is also supported in part 
by a Grant-in-Aid for Scientific Research on Innovative Areas 
``Topological Materials Science" (KAKENHI Grant No. 15H05855) 
from MEXT of Japan.

\makeatletter
\@addtoreset{equation}{section}
\makeatother 

\renewcommand{\theequation}{A\arabic{equation}} 

\appendix

\section{Continuum version of the modified $XY$ model}\label{sec:appA}

In this appendix we show how to derive the Hamiltonian (\ref{Eq:Ham}) from the microscopic lattice model (\ref{Eq:modXY}). The reformulation in terms of a continuum theory is based on the equivalence of the partition function. By definition we have
\bea
Z=\int {\cal D} \vartheta \exp \Big\{\sum_{x,i}\Big[\frac{J}{2} \cos (\nabla_i \vartheta_x)+\frac{J'}{2}\cos (2\nabla_i \vartheta_x)\Big]\Big\}, \nonumber\\
\eea
where $\nabla_i \vartheta_x=\vartheta_x-\vartheta_{x+i}$, the sum over $x$ goes through the whole lattice, the sum over $i$ refers to the neighbors ($i=1,...,4$), and we absorbed the inverse temperature $\beta$ into the couplings $J$ and $J'$. Introducing the notation $\Psi_x=\exp i\vartheta_x$, we have
\bea
Z=\int {\cal D} \vartheta \exp \Big\{\frac{1}{4}\sum_{x,i} (J\Psi_x\Psi^*_{x+i}+J'\Psi^2_x\Psi^{*2}_{x+i}+\cc)\Big\}. \nonumber\\
\eea
Now we introduce a complex field $\psi_x$ via delta functions
\bea
Z&=&\int {\cal D} \vartheta {\cal D} \psi {\cal D} \psi^* \prod_x \delta(\psi_x-\Psi_x)\delta(\psi_x^*-\Psi_x^*) \nonumber\\
&\times& \exp \Big\{\frac{1}{4}\sum_{x,i} (J\psi_x\psi^*_{x+i}+J'\psi^2_x\psi^{*2}_{x+i}+\cc)\Big\}.
\eea
The $\delta$ functions can be represented using a complex auxiliary field $\alpha_x$:
\bea
Z&=&\int {\cal D} \vartheta {\cal D} \psi {\cal D} \psi^* {\cal D}\alpha {\cal D}\alpha^* \exp \Big\{-\frac{1}{2}\sum_{x}\Big[ i(\psi_x-\Psi_x)\alpha_x \nonumber\\
&+& \cc\Big]+\frac{1}{4}\sum_{x,i} (J\psi_x\psi^*_{x+i}+J'\psi^2_x\psi^{*2}_{x+i}+\cc) \Big\}.
\eea
Using that
\bea
\int_0^{2\pi} \frac{d\vartheta}{2\pi} \exp(|\alpha| \cos \vartheta)=I_0(|\alpha|),
\eea
where $I_0$ is the Bessel function, we get
\bea
Z&=&\int {\cal D} \psi {\cal D} \psi^* \exp\Big\{\frac{1}{4}\sum_{x,i} (J\psi_x\psi^*_{x+i}+J'\psi^2_x\psi^{*2}_{x+i}+\cc) \Big\} \nonumber\\
&&\hspace{-0.8cm} \int\!\! {\cal D}\alpha {\cal D}\alpha^*\exp \Big\{-\frac{1}{2}\sum_{x}\Big[ i\psi_x\alpha_x+\cc-2\log I_0(|\alpha_x|)\Big]\Big\}. \nonumber\\
\eea
Using the notation $\rho_x=|\psi_x|^2/2$, we define a potential term $U(\rho_x)$ through the equation
\bea
&&\exp\Big\{-\sum_x \Big[U(\rho_x)+2J|\psi_x|^2+2J'|\psi^2_x|^2\Big]\Big\}= \nonumber\\
&&\hspace{-0.15cm}\int \!\!{\cal D}\alpha {\cal D}\alpha^*\exp \Big\{-\frac{1}{2}\sum_{x}\Big[i \psi_x\alpha_x+\cc-2\log I_0(|\alpha_x|)\Big]\Big\},\nonumber\\
\eea
and as a final step we take the continuum limit. Then
\begin{subequations}
\bea
\hspace{-0.5cm}\sum_{x,i} \psi_x \psi^*_{x+i} &\approx& 4 \sum_x |\psi_x|^2+\int d^2x \psi(x)\Delta \psi^*(x), \\
\hspace{-0.5cm}\sum_{x,i} \psi^2_x \psi^{*2}_{x+i} &\approx& 4 \sum_x |\psi^2_x|^2+\int d^2x \psi^2(x)\Delta \psi^{*2}(x),
\eea
\end{subequations}
and the partition function takes the (continuum) form
\bea
\label{Eq:Zfinal}
Z&=&\int {\cal D}\psi {\cal D}\psi^* \exp \Big\{ -\int d^2x \Big[ \frac{J}{2} |\nabla \psi(x)|^2+\frac{J'}{2}|\nabla \psi^2(x)|^2\nonumber\\
&&\hspace{4cm}+U(\rho(x)) \Big] \Big\}.
\eea
Note that we have rescaled the effective potential with the square of the lattice spacing. Using the notation $a=J/2$ and $b=J'/2$, and expanding the potential around its minimum
\bea
U(\rho) \approx \frac{\lambda}{2}(\rho-\rho_0)^2,
\eea
we find that 
Eq.~(\ref{Eq:Zfinal}) is the partition function of a system with the Hamiltonian
\bea
\hspace{-0.4cm}{\cal H}&=&\int_x \left[ a|\nabla \psi|^2 +b|\nabla \psi^2|^2+\frac{\lambda}{2}\big(|\psi|^2/2-\rho_0\big)^2\right],
\eea
which completes the derivation.

\renewcommand{\theequation}{B\arabic{equation}}

\section{Flow equations}\label{sec:appB}

In this appendix, we show how to derive the flow equations (\ref{Eq:lambdarhoflow}) and (\ref{Eq:Zkflow}) of the LPA'. Using the notation
\bea
U_k(\rho)=\frac{\lambda_k}{2}(\rho-\rho_{0,k})^2,
\eea
with the help of Eq.~(\ref{Eq:optreg}), we derive from Eq.~(\ref{Eq:flow1}) that
\bea
k\partial_k U_k = \frac{k^4}{4\pi} \Big[1-\frac{\eta_k}{4}\Big]\Big(\frac{1}{k^2+M_t^2/Z_k^2}+\frac{1}{k^2+M_l^2/Z_k^2}\Big), \nonumber\\
\eea
where $M^2_{k,t}$ and $M^2_{k,l}$ are the transversal and longitudinal components of the mass matrix $M^2_k$, respectively,
\bea
M^2_{k,ab}&=&M_{k,t}^2\delta_{at}\delta_{bt}+M_{k,l}^2\delta_{al}\delta_{bl},\nonumber\\
M_{k,t}^2&=&U_k'(\rho), \quad M_{k,l}^2=U_k'(\rho)+2\rho U_k''(\rho),
\eea
and $\eta_k=-\frac{1}{Z_k}\frac{d Z_k}{dk}$ is the anomalous dimension. Expanding the right-hand side with respect to $\rho$, we compare it with the left-hand side and identify the flows $k\partial_k \lambda_k$ and $k\partial_k \rho_{0,k}$ leading to Eqs.~(\ref{Eq:lambdaflow}) and (\ref{Eq:rhoflow}), respectively.

For the flow of $Z_k$ and thus the expression of the anomalous dimension, we let the operator $\delta^2/\delta \psi_j(-p)\delta \psi_i(p)$ act on both sides of Eq.~(\ref{Eq:flow1}). Then we arrive at
\bea
\label{Eq:gamma2flow}
k\partial_k \Gamma_{k,ij}^{(2)}(p,-p)&=&\int_q k\partial_k R_k(q) [\Gamma_k^{(2)}+R_k]^{-1}_{ab}(q)\nonumber\\
&\times&[\Gamma_k^{(2)}+R_k]^{-1}_{cd}(q-p)[\Gamma_k^{(2)}+R_k]^{-1}_{ea}(q)\nonumber\\
&\times&\Gamma_{k,bcj}^{(3)}\Gamma_{k,dei}^{(3)},
\eea
where $\Gamma_k^{(2)}$ and $\Gamma_k^{(3)}$ are the second and third functional derivatives of $\Gamma_k$, respectively,
\bea
\Gamma^{(2)}_{k,ab}(q)&=&(Z_kq^2\delta_{ab}+M_{ab}^2)^{-1}, \\
\Gamma^{(3)}_{k,abc}&=&\lambda_k(\delta_{ab}\psi_c+\delta_{bc}\psi_a+\delta_{ca}\psi_b),
\eea
where we see that the $\Gamma_k^{(3)}$ vertex is momentum independent in the LPA' approximation (\ref{Eq:LPAp}). Note that, in principle, the wave function renormalization factors in the broken phase are different for the longitudinal and transverse components. When deriving the flow of $Z_k$, we take into account only the transverse component. Assuming that $\psi_i=\delta_{il} \psi$ is a homogeneous background, the $tt$ component of Eq.~(\ref{Eq:gamma2flow}) reads
\bea
k\partial_{k}\Gamma_{tt}^{(2)}(p,-p)&=&2\rho\lambda_k^2\int_q k\partial_k R_k(q) \nonumber\\
&\times&\Big[(\Gamma_k^{(2)}+R_k)_{ll}^{-2}(q)(\Gamma_k^{(2)}+R_k)_{tt}^{-1}(q+p) \nonumber\\
&+&(\Gamma_k^{(2)}+R_k)_{tt}^{-2}(q)(\Gamma_k^{(2)}+R_k)_{ll}^{-1}(q+p)\Big].\nonumber\\
\eea
Since $\partial_k R_k(q) \sim \Theta (k^2-q^2)$, the integral is restricted to $0<|q|<k$, and we can substitute $Z_kq^2+R_k(q) \rightarrow Z_kk^2$ in the two-point functions. Then we get
\bea
\label{Eq:gamma2flowb}
k\partial_k &&\!\!\!\!\!\!\!\Gamma_{k,tt}^{(2)}(p,-p)=2\rho\lambda^2_k \int_{|q|<k} f_k(q)(Z_kk^2+M_{k,l}^2)^{-2} \nonumber\\
&\times&[Z_k(q+p)^2+M_{k,t}^2+R_k(p+q)]^{-1}+\{t \leftrightarrow l\}, \nonumber\\
\eea
where $f_k(q)=k[2kZ_k+(k^2-q^2)\partial_k Z_k]$. Now we project both sides of Eq.~(\ref{Eq:gamma2flowb}) onto the ${\cal O}(p^2)$ piece. The left-hand side is simply
\bea
\label{Eq:flowlhs}
\lhs = k\partial_k Z_k(\rho) p^2,
\eea
while for the right-hand side we have
\bea
\label{Eq:flowrhs}
\rhs &=& \frac{2\rho\lambda^2_k}{(Z_kk^2+M_{k,l}^2)}\Bigg[\int_{q_+<|q|<k} \Theta(x>0) \nonumber\\
&\times&\Bigg(\frac{f_k(q)}{Z_k(p^2+2pqx+q^2)+M_{k,t}^2}-\frac{f_k(q)}{Z_kk^2+M_{k,t}^2}\Bigg)\Bigg]  \nonumber\\
&+& \{t \leftrightarrow l\}+{\cal O}(p^3), 
\eea
where $x=\hat{p}\hat{q}$, and $q_+=k-px+{\cal O}(p^2)$. After performing the integral, we compare Eqs.~(\ref{Eq:flowrhs}) with (\ref{Eq:flowlhs}) and arrive at
\bea
k\partial_k Z_k(\rho)p^2 = -\frac{\rho\lambda_k^2k^4Z_k^2}{\pi(Z_kk^2+M_{k,l}^2)^2(Z_kk^2+M_{k,t}^2)^2} p^2, \nonumber\\
\eea
which leads to Eq.~(\ref{Eq:Zkflow}).

\renewcommand{\theequation}{B\arabic{equation}}

\end{document}